# Semi-orthogonal subspaces for value mediate a tradeoff between binding and generalization

W. Jeffrey Johnston[1*+], Justin M. Fine[2*],
Seng Bum Michael Yoo[3,], R. Becket Ebitz[4], and
Benjamin Y. Hayden[2]

1. Center for Theoretical Neuroscience and Mortimer B. Zuckerman Mind, Brain, and Behavior Institute, Columbia University, New York, New York, United States of America
2. Department of Neurosurgery, Baylor College of Medicine, Houston, Texas, United States of America
3. Department of Biomedical Engineering, Sunkyunkwan University, and Center for Neuroscience Imaging Research, Institute of Basic Sciences, Suwon, South Korea, Republic of Korea, 16419
4. Department of Neuroscience, Université de Montréal, Montréal, Quebec, Canada

\* Equal contribution
\+ Correspondence: wjeffreyjohnston@gmail.com

**Funding statement**

This research was supported by a National Institute on Drug Abuse Grant R01 DA038615 (BYH), MH124687 (BYH), NSF 1707398 (WJJ), Simons Foundation 542983SPI (WJJ), Gatsby Charitable Foundation GAT3708 (WJJ)

**Competing interests**

The authors have no competing interests to declare.

**Acknowledgements**

We thank Bill Vinje, who led an excellent summer journal club on the binding problem at UC Berkeley in 2001. We thank Stefano Fusi for useful discussions of previous versions of this manuscript. We thank Maya Wang, Tyler Cash-Padgett, Marc Mancarella, Caleb Strait, Tommy Blanchard, and Brianna Sleezer for assistance with data collection. We also thank Allison Ong and Aleyna Silcott for administrative support.

## ABSTRACT

When choosing between options, we must associate their values with the action needed to select them. We hypothesize that the brain solves this binding problem through neural population subspaces. To test this hypothesis, we examined neuronal responses in five reward-sensitive regions in macaques performing a risky choice task with sequential offers. Surprisingly, in all areas, the neural population encoded the values of offers presented on the left and right in distinct subspaces. We show that the encoding we observe is sufficient to bind the values of the offers to their respective positions in space while preserving abstract value information, which may be important for rapid learning and generalization to novel contexts. Moreover, after both offers have been presented, all areas encode the value of the first and second offers in orthogonal subspaces. In this case as well, the orthogonalization provides binding. Our binding-by-subspace hypothesis makes two novel predictions borne out by the data. First, behavioral errors should correlate with putative spatial (but not temporal) misbinding in the neural representation. Second, the specific representational geometry that we observe across animals also indicates that behavioral errors should increase when offers have low or high values, compared to when they have medium values, even when controlling for value difference. Together, these results support the idea that the brain makes use of semi-orthogonal subspaces to bind features together.

**INTRODUCTION**

Human cognition is remarkably flexible. We can rapidly combine different pieces of information arbitrarily and on the fly. When making economic choices, decision-makers must link the anticipated values of each option with the actions needed to choose the preferred one (Kable and Glimcher, 2009; Samejima et al., 2005; Rangel et al., 2008; Wunderlich et al., 2009). Failure to bind these elements correctly will lead to an erroneous assignment of value to action, and could yield suboptimal choices. The problem of how the brain binds value to action remains unresolved, but is fundamental to the field of neuroeconomics (Cai and Padoa-Schioppa, 2014; Hare et al., 2011; Hayden and Moreno-Bote, 2018; Knudsen and Wallis, 2022; Padoa-Schioppa and Assad, 2008; Rangel et al., 2008; Rosech et al., 2009; Shadlen and Movshon, 1999; Wunderlich et al., 2009; Yin et al., 2019; Stoet and Hommel, 1999). Moreover, the neural mechanisms underlying value-action binding could apply well beyond the neuroeconomics context to general binding between different sensory features.

A means for solving this binding problem is found in the information coding power and flexibility of neural populations (Chung and Abbott, 2021; Ebitz and Hayden, 2021; Elsayed et al., 2016; Mante et al., 2013; Urai et al., 2021; Saxena and Cunningham, 2019; Vyas et al., 2020). The flexibility of population codes is partly attributable to the fact that individual neurons often respond to linear and nonlinear mixtures of multiple task-relevant features (i.e., mixed selectivity; Blanchard et al., 2018; Fusi et al., 2016; Hocker et al., 2021; Raposo et al., 2014; Rigotti et al., 2013). Nonlinear mixed selectivity is conjunctive, where a neuron's response depends not only on the individual values of multiple features, but specifically on different combinations of those values (e.g., a neuron that responds only to red squares rather than to either the color red or the shape square in isolation). This form of selectivity has been shown to support flexible behavior in complex cognitive tasks (Rigotti et al., 2013) as well as to provide more reliable representations than linear selectivity (Barak et al., 2013; Babadi and Sompolinsky, 2014; Litwin-Kumar et al., 2017; Johnston et al., 2020). Further, previous theoretical work has shown that these conjunctive responses can facilitate reliable decoding of multiple stimuli from a single population of neurons (Matthey et al., 2015). Building on this work, we hypothesize that neural systems solve the value-action binding problem by encoding the value of different potential options along semi-orthogonal dimensions of the full population space, which are naturally produced by nonlinear mixed selectivity (Fusi et al., 2016; Parthasarathy et al., 2017). We refer to these lower-dimensional spaces embedded in the full population space as subspaces.

We propose that, in core reward regions, value is bound to other variables – such as spatial position and time of presentation – by use of subspaces (Aoi et al., 2020; Maggi & Humphries, 2022), which we refer to as *binding-by-subspace*. For instance, independent (and therefore non-bound) representations of offer value and position could manifest as a rectangular structure in population space (**Figure 2D**, top). This low-dimensional, factorized coding scheme can unambiguously represent the value and position of a single offer, but cannot unambiguously represent more than one value-position pair at the same time. For two offers represented in a perfectly factorized system, a downstream circuit could separately infer the two offer values and the two positions, but not *which* offer value was at which position. This ambiguity exists even for separate nonlinear representations of value and position (see *Binding by subspace orthogonalization* in **Methods** for a more detailed discussion). One way to break this ambiguity is to represent the value of left and right offers in fully orthogonal subspaces (schematized in **Figure 2D**, bottom). While this does solve the binding problem, there are two drawbacks. First, it requires many highly specific subspaces (i.e., a subspace for every distinguishable spatial position, leading to a combinatorial explosion). Second, each representation of value would be independent of all the others and thus would not be abstract (Bernardi et al., 2020; Johnston & Fusi, 2022). This, in turn, would make transfer learning and generalization to novel situations difficult if not impossible (Dosher & Lu, 2017). Thus, we propose that core reward regions exist in a middle ground: They represent distinct offers in semi-orthogonal subspaces, which simultaneously provide both reliable binding and preserve an abstract representation of value.

Here, we examined the representational geometry of neural population representations in five brain areas associated with reward: the orbitofrontal cortex (OFC 11/13), the pregenual (pgACC 32) and posterior cingulate (PCC 29/31) cortices, the ventromedial prefrontal cortex (vmPFC 14) and the ventral striatum (VS). All regions were recorded in the same risky choice task, involving two risky options separated in space and in time (Strait et al., 2014). First, we investigate spatial binding and find that all regions encode the values of left and right offers in distinct, semi-orthogonal subspaces. To understand how these semi-orthogonal subspaces relate to binding, we develop a mathematical theory that links subspace orthogonality, binding, and abstraction to measurable aspects of the representational geometry within each region. Using this theory, we show that OFC, pgACC, and vmPFC all have representational geometry that is consistent with a near-even tradeoff between misbinding errors and generalization errors, while VS appears specialized toward minimizing generalization errors and an abstract representation of value at the expense of misbinding errors. In contrast, PCC appears specialized for nonlinear and reliably bound representations while sacrificing generalization.

Next, we show that the first and second offer values are represented in orthogonal subspaces, indicating that these regions use subspaces to link offer values to the egocentric time of their presentation. The representational geometry of this combined representation makes specific predictions about behavior, allowing us to more rigorously test the binding hypothesis. First, we find evidence that behavioral error trials are associated with fluctuations in the neural activity that resemble a spatial misbinding error. Crucially, behavioral error trials do not also resemble temporal misbinding errors – consistent with the fact that the subjects eventually make a spatial choice. Finally, we find that stimulus conditions that are less distinct in the representational geometry (high- and low-magnitude offer pairs, as opposed to medium-magnitude pairs) are associated with reduced behavioral accuracy relative to trials that are less confusible in the geometry. Together, these results support the hypothesis that the brain makes use of subspace orthogonalization to solve the value-action binding problem.

## RESULTS

**Subjects show behavior consistent with task understanding in risky choice task**

We analyzed data collected from six rhesus macaques trained to perform a two-option, **asynchronous risky gambling task** that we have used several times in the past (e.g., Strait et al., 2014, **Figure 1**). Behavioral data were consistent with patterns we have previously reported (for the most complete analyses, see especially Farashahi et al., 2018 and 2019). Briefly, on each trial, subjects use a saccade to select one of two risky offers. The two offers are presented in sequence and on different sides of a computer screen (left or right). Each option appears for 400 ms, followed by a blank screen for 600 ms; thus, there is a one-second stimulus onset asynchrony between the two offers. Then, the two offers reappear simultaneously and the subject makes a choice by fixating their preferred offer for 200 ms. Each offer is defined by a probability (0-100%, 1% increments) and stakes (large or medium reward, 0.240 and 0.165 mL juice, respectively). Thus, once the animal selects an offer, the probability of the offer dictates how likely they are to receive any reward and the stakes determine the size of that possible reward. The probabilities and stakes associated with both offers are chosen randomly for each option. The order of presentation (left first vs. right first) is randomized by trial.

All six animals were more likely to choose offers with a higher expected value – i.e., the product between probability and stakes. As in past studies using this task, all subjects were reliably risk-seeking (Heilbronner and Hayden, 2013). For our analyses here, we estimated the subjective value underlying each animal's choices. In particular, while the nominal expected value is the product of the probability and stakes (as mentioned above), the different subjects may differ in how they weigh these two variables. Indeed, the best fitting model we considered incorporates non-linear weightings of probability and stakes, and predicts the animal's choice with 87 - 95% accuracy. The subjective value model is fit independently for each session. In the rest of the manuscript, we use the term *value* to mean the model-derived subjective value for the session from which it was collected.

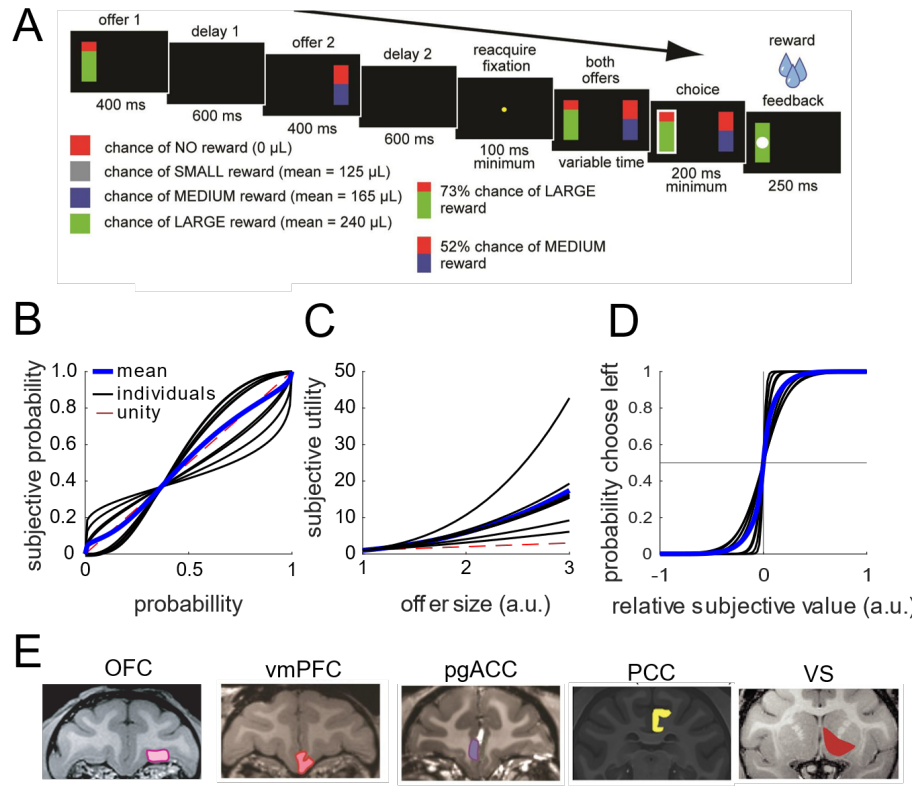

**Figure 1.** Task outline and brain areas. **A.** The risky-choice task is a sequential offer decision-task that we have used in several previous studies (e.g., Strait et al., 2014). In the first 400 ms, subjects see the first offer as a bar presented on either the left or right side. The above shows an example on the left. This offer is followed by a 600 ms delay, a 400 ms offer 2 window, another delay (2) window, and then choice. The full task involved either small, medium or large reward offers. The small reward trials were actually those with safe (guaranteed) offers. We only analyzed the risky choice trials, which were those including the medium and large reward. **B-D.** From left to right, the plots show for all subjects the model fitted (left) subjective probability, (middle) subjective utility, and (right) relative subjective value choice curves. **E.** MRI coronal slices showing the 6 different core reward regions that were analyzed.

**Single neurons have nonlinear responses to different combinations of value and space**

We analyzed the responses of 929 neurons across five brain regions, with two subjects per region (**Figure 1**; n=242 neurons in OFC, n=156 in vmPFC, n=255 in pgACC, n=152 in PCC, and n=124 in VS). We observed diverse responses from neurons in each brain region (**Figure 2A**). These diverse responses also give rise to diverse value-response functions (**Figure 2B**), many of which are consistent with our hypothesis of a nonlinear interaction between value and space (e.g., the example neurons from OFC and pgACC).

To investigate whether these neural responses were best explained by linear or nonlinear interactions between representations of value and space, we fit several linear regression models: (1) a model explaining the neural responses in terms of noise (**Figure 2C**, "noise"); (2) a model with separate terms for value and space (**Figure 2C**, "linear"); (3) a model with terms for value and space as well as a nonlinear interaction between them (**Figure 2C**, "interaction"). For both the linear and interaction models, we fit versions of the model with both linear and spline-based representations for value (see *Single-neuron linear regression model* in **Methods** and **Figure S1** for more details). Then, we compared the fits of these models to the data, through a Bayesian model stacking analysis based on approximate leave-one-out cross validation (see *Comparison between models with parallel and non-parallel subspaces* in **Methods** and Yao et al., 2017; **Figure 2C**). In every brain region, we found substantial proportions of neurons with responses best fit by the nonlinear interaction model (OFC: 12%, PCC: 19%, pgACC: 10%, vmPFC: 12%, VS: 13%). In these neurons, space does not merely shift the value tuning curves of individual neurons, it changes them qualitatively (**Figure 2B** for linear tuning curves; **Figure S1B** for nonlinear tuning curves). Interestingly, these results are inconsistent with previous claims and interpretations present in the literature (including from our own research group, see Strait et al., 2016). Neurons with linear mixed selectivity for value and position could encode both quantities, but linear mixed selectivity alone does not bind a particular value to a particular position. Thus, in our setting, nonlinear mixed selectivity may provide a crucial benefit: it could bind the value of an offer to its position by producing non-parallel subspaces for different spatial positions.

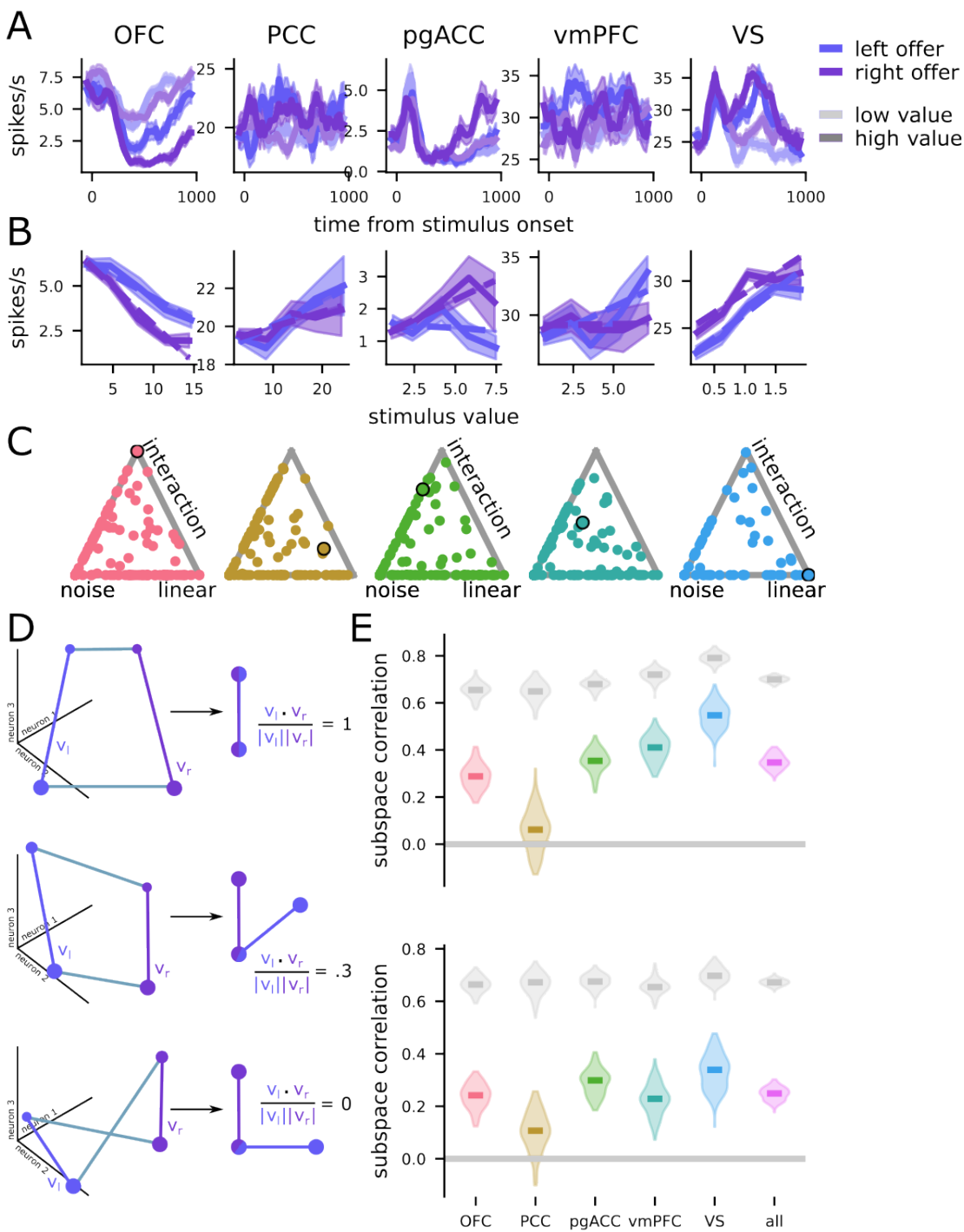

**Figure 2.** Diverse value-response functions produce semi-orthogonal subspaces for value. **A.** The firing rates of example neurons from each region during the offer window, shown for high and low value offers presented on the left or right side (100 ms boxcar filter, shaded area is SEM). **B.** The value-response function for each neuron in **A**. The value-response function fit by the linear regression model with an interaction term is overlaid (dashed lines). We show models with a linear representation of value here (i.e., all the value-response functions are straight lines); see **Figure S1B** for the same

neurons shown with spline-based nonlinear value-response functions. **C.** A simplex showing the weight given to each of the noise-only, linear, and interaction regression models by the Bayesian model stacking analysis. In the simplex, a point in the one of the corners indicates that the neuron is best explained by the corresponding model alone (i.e., that model is given weight 1 and all other models have weight 0). A point in the center of the triangle indicates that the neuron is best explained by an equally weighted combination of all three models. The points corresponding to the example neurons shown in **A** and **B** have dark outlines here. Both the linear and interaction categories include models with both linear and spline representation of value. **D.** Schematic of three different representational geometries that would lead to different subspace correlation results. (top) Two perfectly aligned value vectors $v_l$ and $v_r$ in population space (left) would produce a subspace correlation close to 1 (right). (middle) Two partially aligned value vectors $v_l$ and $v_r$ in would produce a subspace correlation between 0 and 1 (note there is an additional possibility: partially aligned but negatively correlated subspaces; not schematized). (bottom) Two unaligned value vectors $v_l$ and $v_r$ would produce a subspace correlation close to 0. **E.** Subspace correlations computed across the full pseudopopulation recorded from each region for the offer presentation window (top) and the delay period (bottom). The rightmost "all" datapoint is the full recorded pseudopopulation, combined across all regions. The gray point is the subspace correlation expected if the left- and right value subspaces were aligned and corrupted only due to noise.

**Neural populations use separable but partially overlapping subspaces for different offers**

What does the single neuron result in the previous section mean at the population level? To quantify the degree of overlap between the left- and right-value subspaces, we use the coefficients from the regression model (see above), which define a value-encoding subspace for the left ($v_l$) and right ($v_r$) sides (**Figure 2D**; and see **Figure S1** for a replication of these results when the subspaces are not constrained to be vectors).

A correlation between left and right vectors that is close to 1 (or, technically speaking, that is close to the noise ceiling, shown in gray in **Figure 2E**) indicates primarily linear interactions between the offer value and position encodings (**Figure 2D**, top; the two value vectors $v_l$ and $v_r$ are parallel to each other). As explained above, such a result would indicate that the coding scheme cannot implement subspace binding. Conversely, subspace correlations substantially less than the noise ceiling indicate nonlinear interactions between representations of offer value and position, and are consistent with our hypothesis of subspace binding. A value greater than zero (**Figure 2D**, middle) would mean that there is a mixture of both linearly and nonlinearly interacting value and position representations. The intermediate level of subspace correlation (i.e., between 1 and 0) supports binding and has an additional benefit over the zero correlation scheme: it allows for the generalization of the value code across offer positions and epochs (Dosher & Lu, 2017; Bernardi et al., 2020; Johnston & Fusi, 2022). Other possibilities that our analysis approach could detect would include no subspace correlation (**Figure 2D**, bottom; the two value vectors $v_l$ and $v_r$ are fully orthogonal), or negative subspace correlation (not schematized).

We first computed the subspace correlation between left and right offers during the 400 ms offer presentation window. We find that in all five brain regions, the subspaces for left and right offers are significantly less correlated than the noise ceiling ($p < 0.001$ in all cases, **Figure 2E**, top; and see *Computing population encoding subspaces for space and value* in **Methods** for reliability control, cf. Kimmel et al., 2020). Subspace correlations in every region except PCC are also significantly greater than zero ($p < 0.001$, **Figure 2E**, top). We found similar results when repeating the above comparison for the delay period (all $p < 0.001$, **Figure 2E**, bottom). These differences can be used to solve the value-space binding problem, and we provide a theoretical prediction for the reliability of that solution below  (**Figure 3**).

**Semi-orthogonal subspaces offer both binding and generalization**

Next, we show that an intermediate (between orthogonal and collinear subspaces) code is able to support binding. Semi-orthogonal representations of the same stimulus feature in different contexts have been observed in many other studies (e.g., Elsayed et al., 2016; Flesch et al., 2022; Tang et al., 2020; Yoo and Hayden, 2020), similar to the semi-orthogonal representations of left- and right-offer value observed here. Why would this kind of representation be favored? Here, we show that, in the noiseless case, a linear decoder can recover the two correct value-position pairs from a linear, additive representation constructed using any subspace correlation value less than 1 (or greater than -1). This follows from the fact that the linear encoding matrix for the two stimuli will have an inverse so long as the vectors that encode the two stimuli are not perfectly parallel (see *Subspace binding for continuous variables* in **Methods** for more details).

Next, we include noise and show that the total error in the decoded offer values strictly decreases as the absolute value of subspace correlation decreases. This means that more orthogonal subspaces will lead to more reliable decoding, and that a subspace correlation of zero will provide the most reliable representation of the two offer values (**Figure 3A**). However, we find subspace correlations that are significantly greater than zero for all brain regions that we investigated (**Figure 2E**). Could these non-zero subspace correlations confer a different advantage? Following previous work (Dosher & Lu, 2017; Bernardi et al., 2020), we posit that non-zero subspace correlations provide an abstract representation of value, which facilitates generalization across contexts and rapid learning.

To investigate generalization in this context, and concretely link our analytic theory to the neural data, we binarize the continuous offer value variable into two categories: high and low (see *Binarizing value* in **Methods** for details). So, we consider representations that consist of four points; for space, these are high value offers presented on the left, high value offers presented on the right, and similarly for low value offers. We also consider an analogous set of four points divided across different presentation times, replacing left and right with offer 1 and offer 2.

Then, we develop a theory that treats the geometry of these kinds of discrete representations as resulting from two components: First, a rectangular scaffold, where one axis of the rectangle corresponds to offer value and the other corresponds to offer position (**Figure 3B**, yellow lines). Second, high-dimensional perturbations applied to that scaffold for each of the four points (**Figure 3B**, dark purple lines). We refer to the length of the rectangular value axis as the linear distance (**Figure 3B**, $d_L$) and the length of the high-dimensional perturbations as the nonlinear distance (**Figure 3B**, $d_N$). The linear and nonlinear distances are directly related to subspace correlation from before (**Figure 3C** and see *Relating subspace correlation to linear and nonlinear distance* in **Methods**). In particular, a large linear distance and small nonlinear distance implies a high subspace correlation (**Figure 3C**, left); similar linear and nonlinear distances imply a moderate subspace correlation (**Figure 3C**, center); and a large nonlinear distance and small linear distance implies a low subspace correlation (**Figure 3C** right).

Next, we relate these linear and nonlinear distances to the likelihood of different kinds of errors made by a decoder attempting to reconstruct a set of stimuli represented in the code. First, we ask when two stimuli (i.e., two value-position pairs) are being represented, how likely is a decoder to make any kind of error? We show that this overall error rate depends on both the linear and nonlinear distances – and so depends indirectly on the subspace correlation (**Figure 3D**, left). This total error rate includes the rate of misbinding errors (**Figure 3D**, middle), where the value-position associations are mixed up and which are the focus of this study. Interestingly, our theory shows that the misbinding error rate depends only on the nonlinear distance; it does not depend on the linear distance. As a consequence of this, while representations with low subspace correlation will tend to have low misbinding error rates, representation with high subspace correlation can also have low misbinding error rates so long as the nonlinear distance is large relative to the noise. We also show that the misbinding error rate is in tension with another important feature of the code: the ability to generalize a representation of value learned at one spatial location to a second spatial location. We refer to the error rate at the second, untrained location as the generalization error rate (**Figure 3D**, right; this is similar to the cross-condition generalization performance used in Bernardi et al., 2021). Our theory shows that the generalization error rate decreases with linear distance, but increases with nonlinear distance. Thus, increasing the nonlinear distance to reduce misbinding errors will come at a cost: An increased rate of generalization errors. Both error rates also depend on the properties of the stimulus set, such as the number of features (**Figure 3E**) and number of discrete values (**Figure 3F**) that each feature can take on. While misbinding and generalization error rate are always in tension with each other (i.e., the misbinding error rate is lowest for low subspace correlations and the generalization error rate is lowest for high subspace correlations), our theory reveals that even for conditions beyond those of our experiment (with more features or values) there exists a region with low rates of both kinds of error, but that this region becomes smaller as either the number of features (**Figure 3E**) or number of values for each feature (**Figure 3F**) increases. Because the subspace correlations we measured were between zero and one (**Figure 2E**), we hypothesize that the neural representations will be consistent with both low binding error rates and low generalization error rates.

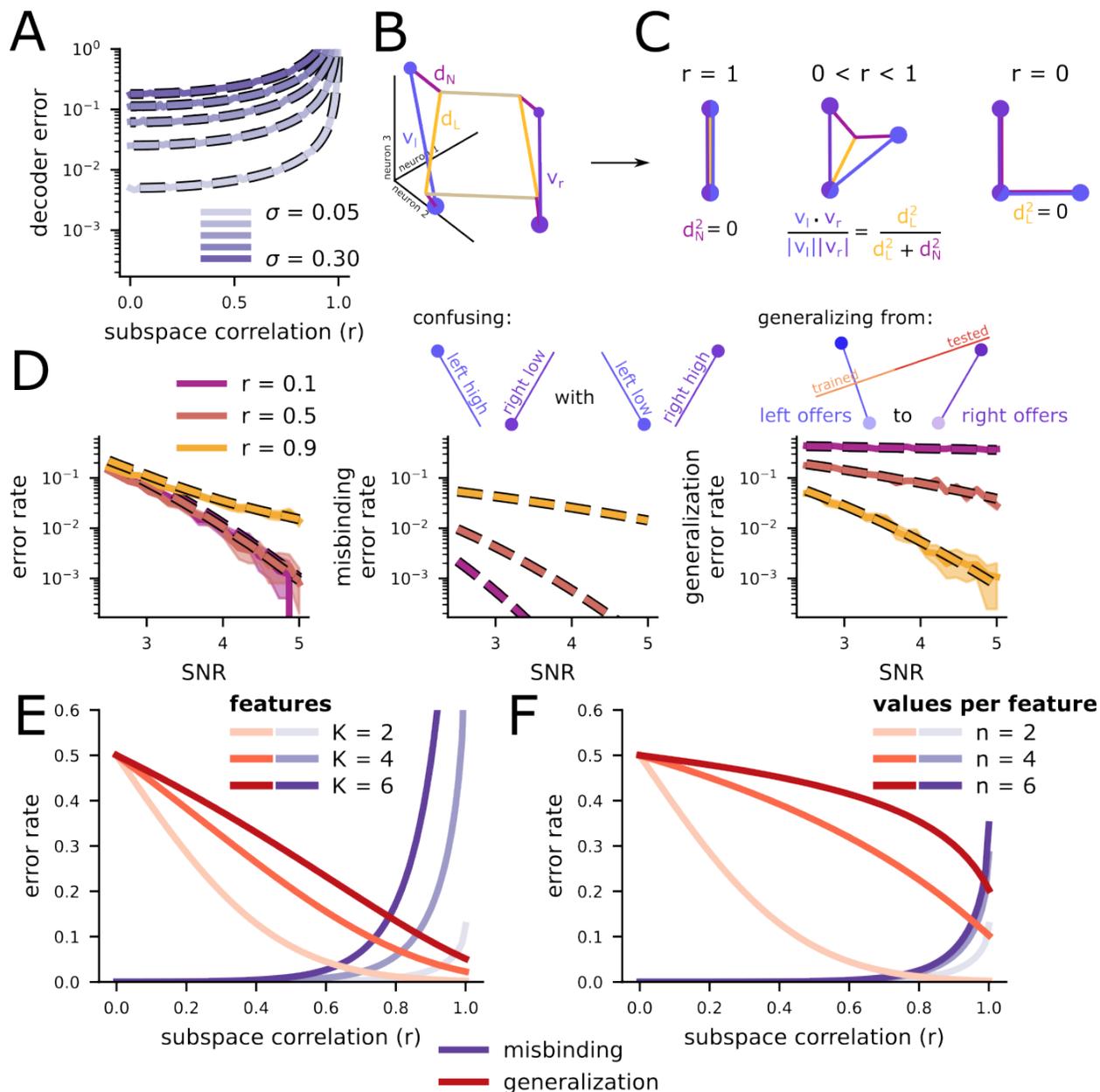

**Figure 3.** Subspace correlation mediates a tradeoff between the reliability of binding and generalization. **A.** The overall error rate for continuous representations of multiple offer values at different noise levels and subspace correlations. The dashed line is our analytic theory. **B.** Schematic of the geometric decomposition into linear and nonlinear distances in the discrete case. **C.** The relative magnitude of linear and nonlinear distance determine the subspace correlation used earlier. **D.** The overall error rate (left), misbinding error rate (middle), and generalization error rate (right) for a discrete code in conditions analogous to the experiment (with two features, K = 2, that each take on two values, n = 2). The dashed line shows the analytic theory, compared to simulations in the left and right plots. Schematics of misbinding errors (middle) and generalization errors (right) are shown on top of the respective plots. **E.** The dependence of the misbinding (purple) and generalization (red) error rates on the number of features in the stimulus set. **F.** The same as **E** but changing the number of values that each feature takes on.

**Value representations are consistent with low binding and generalization error rates**

Now, we apply the theoretical framework developed in the previous section to our experimental data. First, we estimate the distances between all pairs of the four points using an unbiased distance metric (*crossnobis distance*, CITE), which avoids the positive bias common to many other methods of distance estimation. Then, we decompose the resulting distance matrix into linear and nonlinear components, as defined above (and see *Linear-nonlinear data decomposition* in **Methods** as well as **Figure S3** for the decomposition applied to synthetic data). We apply this analysis separately to the left- and right-value code (averaging over the first and second offers) as well as the first and second offer-value code (averaging over left and right offers). For the space-value representation, we find significant nonlinear distances for all individual brain regions except VS and significant linear distances for all regions (one-sided bootstrap test, $p < 0.05$; **Figure 4**, left). We find significant nonlinear and linear distances when combining neurons across the whole population (**Figure 4A**, left, "all"). For the time-value code, we do not find significant nonlinear distances for any region (one-sided bootstrap test, $p > 0.05$), but do find significant linear distances for every region as well as the combined population (one-sided bootstrap test, $p < 0.05$; **Figure 4A**, right).

We now combine these nonlinear distance estimates with the theory developed in the previous section to produce a predicted rate of binding errors. The theory shows that all regions with a nonlinear distance greater than zero will also have a binding error rate that is less than chance. Thus, for the space-value representation, all regions except VS have predicted binding error rates that are less than chance (**Figure 4B**, left). This means that, despite the lack of perfect subspace orthogonality, a downstream neural population would still be able to leverage the distinct subspaces to reliably bind offer value to location. However, this is not the case for the time-value representation (**Figure 4B**, right), where we do not find significant evidence for reliable binding. This indicates that the first and second offer are represented similarly to each other at the time of their presentation. However, there is an important subtlety here. While the currently presented offer is represented similarly whether it is the first or second offer presented in a trial, we show below that the current offer is represented in a distinct subspace from the remembered offer. In particular, we find evidence that time does play a role in binding, but that offers are bound to their egocentric time of presentation (i.e., current or past) rather than their allocentric time of presentation (i.e., first or second). This is in keeping with previous work that shows that neural population representations tend to rotate across time (Libby & Buschman, 2021; Pu et al., 2022).

Finally, we also use the estimated distances to produce a prediction for the generalization error rate of both the value-space and value-time codes. Due to the significant linear components found for every brain region across both the space-value and time-value codes, as well as the only moderate nonlinear distances, the predicted generalization error rates are below chance for every brain region in both codes (**Figure 4C**, open circles). Further, in this case, we can also compute the empirical generalization error rate of a linear classifier – that is, a classifier trained to decode the value of left offers then tested only on right offers. This empirical generalization performance is nearly identical to the predicted performance (**Figure 4C**, outlined circles, and see **Figure S2** for the standard decoding performance). This agreement indicates that our theory captures the aspects of the representational geometry that are relevant to decoding performance. As a consequence, we believe it lends indirect support to the validity of the misbinding error rate also predicted by our theory (**Figure 4B**). Together, this analysis framework identifies each of the regions that we studied here on the binding and generalization error rate plane (**Figure 4D**). Interestingly, most regions do not specialize toward either highly nonlinear or highly linear representations (thereby minimizing only one type of error); instead, they exist at a midpoint, where they simultaneously have large linear and nonlinear distances – and thereby reducing both forms of errors. There are two exceptions to this conclusion, though: PCC appears to specialize for nonlinear representations while VS appears to specialize for linear representations. This supports the idea that VS is a key region for the representation of abstract value (Roesch et al., 2009; Strait et al., 2015). It also supports the idea that PCC may be specialized for spatial representations (Sutherland & Hoesing, 1993; Dean & Platt, 2006) – and we explore how this connects to decision-making below.

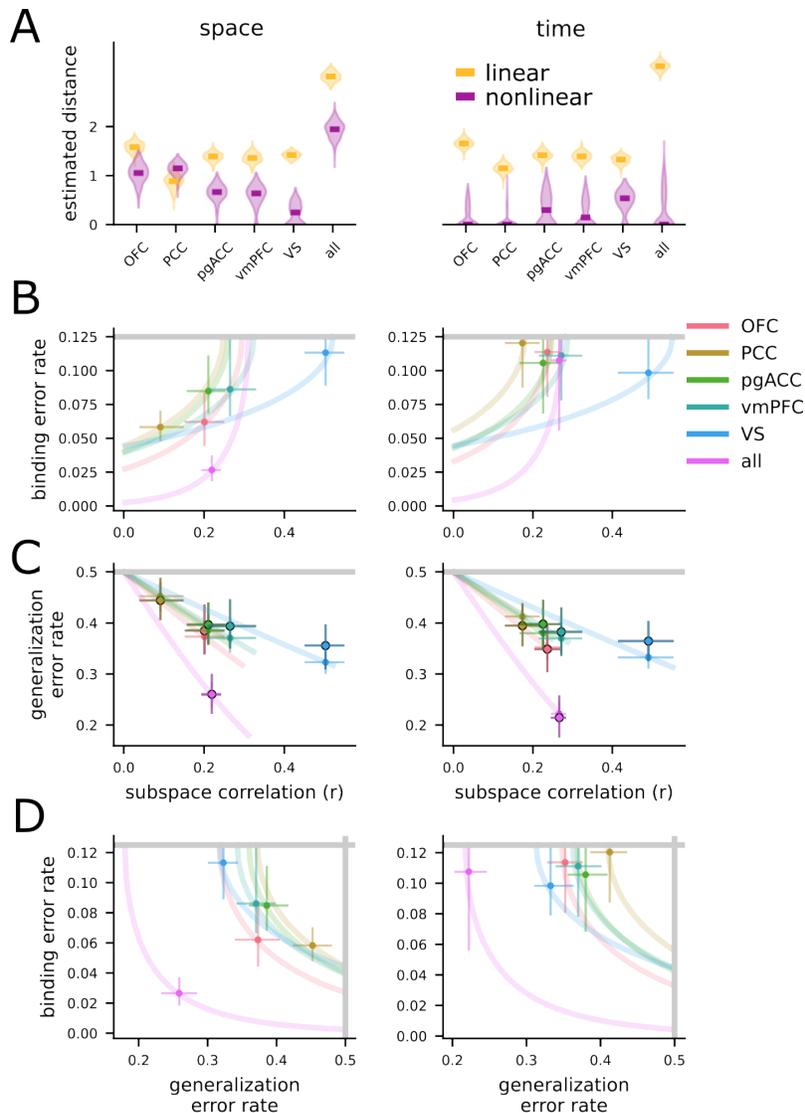

**Figure 4.** The theory predicts misbinding and generalization rates for each region. **A.** The nonlinear and linear distances estimated for the left and right (left) as well as first and second (right) offer value representations within each brain region. The violin plot shows the distribution of bootstrap resamples for the distance estimation procedure. The left and right plots have the same convention for the rest of the figure. **B.** The predicted binding error rate as a function of subspace correlation for each region, derived from the distance estimates in **A**. The gray line shows the chance level of binding errors. The extended color line shows the predicted binding rate for codes with the same total power but different tradeoffs between linear and nonlinear power. **C.** The predicted generalization error rate as a function of subspace correlation for each region, derived from the distance estimates in **A** (for the open circles) and computed empirically with a linear decoder (for the outlined circles). The gray line is the chance level. **D.** Each region shown on the plane defined by the generalization and binding error rates, derived from the distance estimates in **A**. The gray lines are the chance levels for each of the error types.

**The remembered offer value is stored in an orthogonal subspace to the current offer value**

Above, we show that offers are not bound to their allocentric time of presentation – that is, an offer value is not encoded significantly differently if it is the first rather than the second offer (**Figure 4A**, right). Here, we ask whether offers are bound to their egocentric time of presentation instead. That is, we ask whether offer values are encoded differently if they are currently being presented relative to when they are being remembered from a previous presentation (**Figure 5A**).

We analyzed the simultaneous representation of both offers during the presentation of the second offer and the delay period following it. We find that the value of the remembered (i.e., first) offer can be decoded during this period in all individual brain regions except VS as well as the combined population of all neurons (one-sided bootstrap test, $p < .05$). Next, we repeat our subspace correlation analysis using a linear model of neural responses that now has two sets of terms: one for representing the value and side of the past offer (offer 1) and one for representing the value and side of the current offer (offer 2). We find that subspace correlations between the past and current offers are orthogonal in all brain regions ($p < .001$ relative to the noise ceiling and $p > 0.05$ relative to zero; **Figure 5B**). Given the benefits of semi-orthogonality discussed above, it is surprising that we find full orthogonality instead of semi-orthogonality in this context. Further work could explore whether we see consequences of this in the learning and generalization behavior of the animal. Alternatively, this orthogonality could be important to developing a reliable representation of the difference between the two offer values from a particular trial when the animal is getting ready to choose one of the two offers.

**The simultaneous representation of both offers predicts choice behavior**

Here, we work to understand how the representational geometry of both offers correlates with the behavior of our subjects. To do this, we discretized value into high and low as before, which gives rise to eight distinct trial conditions. Then, we estimated the unbiased distance between each pair of conditions in the full population space and arrived at a representational dissimilarity matrix for the eight conditions (**Figure 5C**). If two conditions are close together in neural population space, they are more likely to be confused by a decoder than if they are far apart.

In this setting, there are three distances that are particularly relevant to binding. First, a temporal binding error would occur when trials with a high value offer on the left followed by a low value offer on the right (high left then low right) are confused with low right then high left trials (**Figure 5D**, left). Second, a spatial binding error would occur when high left then low right trials are confused with low left then high right trials (**Figure 5D**, middle). Third, a combined spatial and temporal binding error occurs when high left then low right trials are confused with low right then high left trials (**Figure 5D**, right). This distance-based analysis also reveals that PCC has a distinct geometry from OFC, pgACC, vmPFC, and VS. In particular, PCC has significant distance between spatial misbinding conditions, but not temporal misbinding conditions (**Figure 5E**, PCC); while OFC, pgACC, and vmPFC have significant distance between temporal but not spatial misbinding conditions (**Figure 5E**, OFC, pgACC, vmPFC). VS and the combined population both have significant distance between spatial and temporal misbinding conditions, but significantly larger distance for temporal relative to spatial misbinding conditions (**Figure 5E**, VS and "all"), while PCC has the opposite. We find that the distance between conditions that would be confused in a spatial-temporal binding error are greater than zero in all regions (one-sided bootstrap test, p < .001). Since the subjects in our task eventually do make a spatial choice (they choose by shifting their eyes to the left or right), we predict that trial-to-trial fluctuations along this spatial misbinding dimension may be particularly important for behavior.

How can we test this prediction? To do so, we turn to a trial-by-trial analysis implemented at the experimental session level. To perform population analyses, we restrict our analysis only to sessions with 10 or more neurons, yielding 12 sessions (6 from OFC, 2 from PCC, and 4 from pgACC). We train a decoder on neural activity following the presentation of the second offer to read out whether the value of the left offer or right offer is higher. We train the decoder using only trials where the animal eventually chooses the higher offer. Then, we compute the average projection toward the correct classification from the decision boundary learned by the decoder on held-out trials. First, we compute this distance on held-out trials with optimal choices (**Figure 5F**, left, x-axis). Then, we compute this distance on held-out trials with non-optimal choices (**Figure 5F**, left, y-axis). The decoders from OFC and PCC have significantly worse average projections on trials with non-optimal choices relative to trials with optimal choices (t-test on the difference in average across sessions, OFC: p = 0.04; PCC: p < 0.001; pgACC: p = 0.36; **Figure 5F**, left). However, the same relationship does not hold for either region (or for pgACC) when we train a decoder to readout whether the first or second offer is higher (t-test on the difference in average across sessions, OFC: p = 0.24; PCC: p = 0.33; pgACC: p = 0.64; **Figure 5F**, right). Thus, not only does the representation change on trials with a suboptimal choice, but it changes specifically toward a spatially (and not temporally) misbound representation. While we only had a handful of sessions for PCC, this result also provides intriguing evidence for a specialized role of PCC in binding offer value to spatial position.

Finally, we consider trials in which both offers are either low or high. In the distance analysis above, trials with a low (high) offer presented on the left followed by a low (high) offer presented on the right are indistinguishable from the representation of a low (high) offer presented on the right followed by a low (high) offer presented on the left (**Figure 5G**) – in our setting, low (and high) offers are not equivalent and take on many different values. This indicates the trials with both low or both high offers do not strongly bind the specific offer value to time or position. Thus, we predict that the animals may be more likely to make suboptimal choices when both offers are low or high relative to when one offer is high and the other offer is low. So, we re-analyze the behavior of the animals for low-low and high-high trials compared to low-high or high-low trials and show that, on average, more suboptimal choices occur for both the low-low and high-high conditions, even when the difference in expected value is the same (**Figure 5H**, left). This behavioral finding is consistent with the significantly increased representational distance for high-low and low-high conditions relative to low-low and high-high conditions (bootstrap test, p < .001; **Figure 5H**, right).

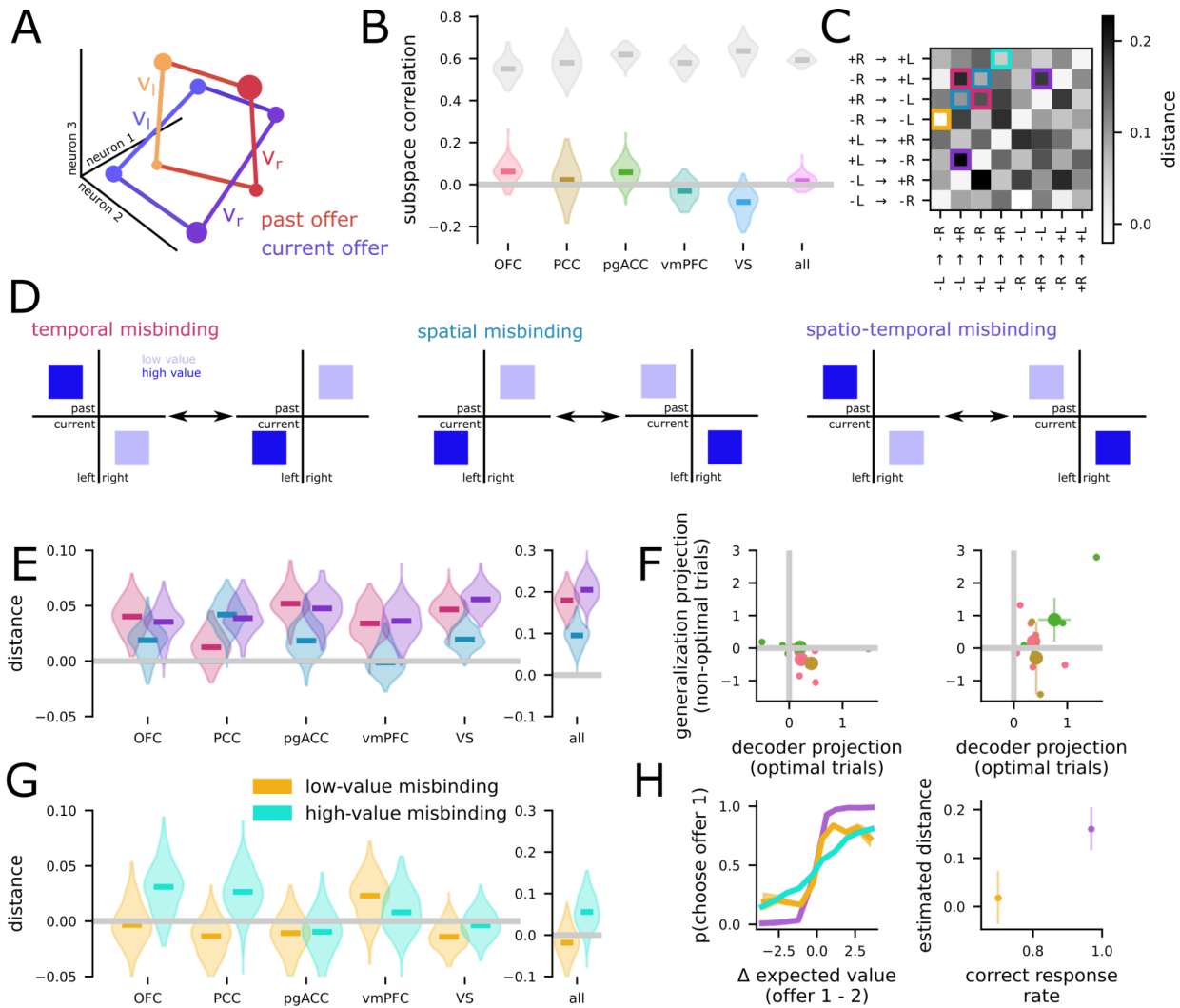

**Figure 5.** The representation of past and current offers predicts key elements of animal behavior. **A**. Schematic of an encoding of the current and past offers together in population space, on semi-orthogonal manifolds. **B**. The subspace correlation between representations of the current and past offer, computed during the presentation of the second offer and following delay period. **C**. Representational dissimilarity matrix for the eight experimental conditions present in the experimental task. To interpret the column and row labels: "+R" indicates a high offer presented on the right side, while "-L" indicates a low offer presented on the left side. The past offer is shown on the left of the arrow while the current offer is on the right of the arrow. **D**. Schematic of the different forms of misbinding error that can occur on this task. (left) Two conditions confused in a temporal misbinding error, where the side of the high offer is the same but its time of presentation is different. (middle) Two conditions confused in a spatial misbinding error, where the side of the high offer is different but its time of presentation is the same. (right) Two conditions confused in a spatio-temporal misbinding error, where the side and presentation of the high offer both differ. **E**. The estimated distances between conditions that – if confused – would constitute temporal, spatial, or spatio-temporal misbinding errors. **F**. The decoding performance of a classifier trained to readout

whether the left or right offer was higher (left) and whether the first or second offer was higher (right). In both cases, the decoder is trained on trials with correct responses (cross-validated performance on correct trials is shown on the x-axis) and tested on trials with incorrect responses. **G**. The estimated distances between conditions with either both high or both low offers. **H**. (left) The behavioral performance of the animal on trials with only high, only low, or a mixture of high and low offers. (right) The estimated distances plotted against the rate of correct responses in each condition.

# DISCUSSION

We asked how the values of two different offers are represented in neural populations from five core cortical reward areas. We find that the different offer values are represented in distinct subspaces, corresponding to both their spatial position and egocentric time of presentation. Further, we show that all of the regions except PCC retain an abstract representation of value across the two different positions investigated here, which is crucial for generalization and rapid learning. Further, we show that neural responses that resemble spatial misbinding are associated with behavioral choice errors, as are experimental conditions associated with geometrical ambiguity. Thus, the subspace features we observe may contribute directly to binding. Together, these results suggest a novel solution to a classic problem in neuroeconomics, one that is grounded in the remarkable coding properties of neuronal populations. Moreover, they raise the possibility that subspace orthogonalization may be a general solution to other important binding problems, such as the perceptual binding problem (Treisman & Gelade, 1980; Roelfsema, 2023).

Since reliable binding is crucial to coherent behavior – and we argue that reliable binding is achieved through orthogonal subspaces for distinct stimuli – why are the subspaces that we observe experimentally semi-orthogonal, rather than fully orthogonal? Our theory shows that both orthogonal and semi-orthogonal subspaces allow for the binding of value to space.

However, the semi-orthogonal subspaces that we find here are simultaneously aligned enough to permit generalization of value across offer positions (Bernardi et al., 2020). A code that permits generalization of value across spatial positions has several benefits: First, it allows the mapping between visual stimulus and value to be learned simultaneously for offers presented at multiple different positions, thus allowing learning from fewer trials. Second, it permits the animal to generalize the mapping it already knows to novel offer positions, thus allowing for significant transfer to novel situations (Dosher & Lu, 2017). In addition, previous theoretical work has shown that semi-orthogonal representations of colors presented at different spatial positions can explain the pattern of errors observed in a common human working memory task (Matthey et al., 2015). While previous work has demonstrated a relationship between representational geometry and generalization to novel samples from the same distribution (Barak et al., 2013; Sorscher et al., 2022), our work provides a link between this geometry and generalization to samples from a systematically different distribution. The generalizability of value is also in line with results from other studies examining perceptual learning of magnitude and ordinal rank, which show generalizability of other scalar parameters (Sheahan et al., 2021).

The result that both low misbinding error rates and above-chance generalization can co-exist is important given wide ranging work about the relative benefits of low- versus high-dimensional neural geometries (Cueva et al., 2020; Flesch et al., 2022; Gallego et al., 2017; Jazayeri and Ostojic, 2021). On one hand, the high-dimensional geometries that arise from nonlinear mixed selectivity have been shown to support flexible decision making on complex cognitive tasks (Rigotti et al., 2013). In contrast, the relatively low-dimensional geometries that often arise from abstract, or factorized, representations of different task variables have been shown to support generalization across contexts and rapid learning (Bernardi et al., 2020;

Johnston & Fusi, 2023; Cueva et al., 2020; Flesch et al., 2022; Gallego et al., 2017; Sohn et al., 2019). These distinct benefits have driven debates about how dimensionality should be interpreted with respect to linking neural coding and behavior (Jazayeri and Ostojic, 2021). Here, we contribute to this work by providing a theory that shows an explicit tension between these benefits – and that shows when the benefits of both forms of representation can be achieved through an intermediate representational geometry. This kind of intermediate geometry has been shown in both mice and monkeys (Bernardi et al., 2020; Nogueira et al., 2021; Boyle et al., 2022; Yoo and Hayden, 2020; Fine et al., 2023). Further, the analytic decomposition into linear and nonlinear distances as well as the theoretical link between these distances and behavioral errors could be applied to understand how the representational geometry – and associated behavior – changes in different contexts.

While we have focused on representations of offer value that are tied to specific spatial positions (i.e., offers presented on the left and right) and egocentric presentation times (i.e., current and past offers), this representation can coexist with representations that depend on or are more closely tied to the animal's eventual choice. In particular, some previous studies have found increased representation of the eventually chosen offer (Gore et al., 2023; Strait et al., 2014) or representations of the difference in value between the two offers, rather than the value of each offer separately (Strait et al., 2015). The existence of these other representations alongside the spatially bound representations that we study here would not change the main conclusions of our study. Further, the spatially bound representations that we study may underlie the formation of these other representations. For instance, a spatially bound representation of the first offer value would be useful for developing a representation of the difference in offer values across the two spatial positions. From an analysis perspective, not including terms for chosen and unchosen offers makes our model less specified rather than incorrectly specified. While this would bias our results given strong choice encodings, the bias would manifest equally in both the measured subspace correlation value and the noise threshold – thus, the qualitative results reported here would not change. Taking our results alongside the conclusions from previous studies suggests that neurons in core value regions represent offer values such that the particular values are bound to particular spatial positions – alongside a nonlinear modulation by the animal's choice.

We and others have noted the encoding of spatial information in single neurons in value-sensitive regions of the brain (Strait et al., 2016; Yoo et al., 2018; Roesch and Schoenbaum, 2006; Feierstein et al., 2006; Tsujimoto et al., 2009; Luk and Wallis, 2013). However, the functional meaning of such signals has been unclear. We have previously conjectured that these spatial signals modulate the encoding of task variables like value through gain-like changes that do not alter tuning (Strait et al., 2016; Yoo et al., 2018; Hayden and Moreno-Bote, 2018). Our present results indicate that our previous hypotheses were partially incorrect. Instead, it also appears that space alters the tuning profile of value-sensitive neurons. Further, spatial position of the offers and the animal's eventual action were confounded in our task, while recent work shows that the contingency between spatial position and action can restructure tuning in frontal cortex (Jonikaitis & Zhu, 2023). Neurons that exhibit a shift in preferred tuning (for either action or spatial position) are expected when they interact with gain modulation neurons in networks with diverse and recurrent connectivity (Salinas and Sejnowski, 2001; Zhang and Abbott, 2000).

While the task used in this work is purely visual, subspace binding can apply to multi-sensory stimuli as well. In this case, however, there is one additional problem that must be solved before the stimuli can be fully integrated. In particular, the distinct representations of the stimuli developed in distinct sensory pathways must be integrated with each other in a process referred to as representation assignment (Johnston & Freedman, 2023). Representation assignment can be solved through redundant stimulus information across sensory systems, such as independent estimates of location from both the visual and auditory systems. Then, the integrated stimuli can be represented as described here: In distinct subspaces depending on their time and location of presentation.

The question of how decision-related information such as value is transformed into action is one of the major ones in the field of neuroeconomics (Kable and Glimcher, 2007; Krabijch et al., 2010; Knudsen and Wallis, 2022; Rangel et al., 2008; Padoa-Schioppa and Assad, 2006). Typical theories hold that values are represented in an abstract value space that is conceptually and functionally distinct from the action space needed to implement the choice of the action. As such, there is a question of how value links up with action. This problem is often reified in neuroanatomy - some regions are assumed to be pure value regions while other, presumably anatomically downstream regions, are assumed to have action signals (Rangel et al., 2008 Kable and Glimcher, 2009; Padoa-Schioppa, 2011). Our results here suggest a somewhat different conclusion. They suggest that the same neural code provides an abstract representation of value as well as representations of value that are bound to particular spatial positions. Further, we do not find that the neurons supporting these different aspects (abstract as opposed to spatially bound) are divided into distinct subpopulations (**Figure S4**). This argues against the idea that there is an anatomical distinction between value and action frames (Fine and Hayden, 2022; Hayden and Niv, 2021). Our findings suggest that this mechanism for binding through semi-orthogonal subspaces is used throughout the cortical reward system. Together, these results highlight the potential value of functional specialization through population representation, rather than through modular architecture, for solving long-standing problems in neuroeconomics

(Ebitz and Hayden, 2021; Urai et al., 2021; Saxena and Cunningham, 2019).

# METHODS

**Surgical procedures:** All procedures were approved by either the University Committee on Animal Resources at the University of Rochester or the IACUC at the University of Minnesota. Animal procedures were also designed and conducted in compliance with the Public Health Service's Guide for the Care and Use of Animals. All surgery was performed under anesthesia. Male rhesus macaques (*Macaca mulatta*) served as subjects. A small prosthesis was used to maintain stability. Animals were habituated to laboratory conditions and then trained to perform oculomotor tasks for liquid rewards. We placed a Cilux recording chamber (Crist Instruments) over the area of interest. We verified positioning by magnetic resonance imaging with the aid of a Brainsight system (Rogue Research). Animals received appropriate analgesics and antibiotics after all procedures. Throughout both behavioral and physiological recording sessions, we kept the chamber clean with regular antibiotic washes and sealed them with sterile caps.

**Recording sites:** We approached our brain regions through standard recording grids (Crist Instruments) guided by a micromanipulator (NAN Instruments). All recording sites were selected based on the boundaries given in the Paxinos atlas (Paxinos et al., 2008). In all cases we sampled evenly across the regions. Neuronal recordings in OFC were collected from subjects P and S; recordings in rOFC were collected from subjects V and P; recordings in vmPFC were collected from subjects B and H; recordings in pgACC were collected from subject B and V; recordings from PCC were collected from subject P and S; and recording in VS were collected from subject B and C.

We defined **OFC 11/13** as lying within the coronal planes situated between 28.65 and 42.15 mm rostral to the interaural plane, the horizontal planes situated between 3 and 9.5 mm from the brain's ventral surface, and the sagittal planes between 3 and 14 mm from the medial wall. The coordinates correspond to both areas 11 and 13 in Paxinos et al. (2008). We used the same criteria in a different dataset (Blanchard et al., 2015).

We defined **vmPFC 14** as lying within the coronal planes situated between 29 and 44 mm rostral to the interaural plane, the horizontal planes situated between 0 and 9 mm from the brain's ventral surface, and the sagittal planes between 0 and 8 mm from the medial wall. These coordinates correspond to area 14m in Paxinos et al. (2008). This dataset was used in Strait et al., 2014 and 2016.

We defined **pgACC 32** as lying within the coronal planes situated between 30.90 and 40.10 mm rostral to the interaural plane, the horizontal planes situated between 7.30 and 15.50 mm from the brain's dorsal surface, and the sagittal planes between 0 and 4.5 mm from the medial wall. Our recordings were made from central regions within these zones, which correspond to area 32 in Paxinos et al. (2008). Note that the term area 32 is sometimes used more broadly than we use it here, and in those studies encompasses the dorsal anterior cingulate cortex; we believe that that region, which is not studied here, should be called area 24 (Heilbronner and Hayden, 2016).

We defined **PCC 29/31** as lying within the coronal planes situated between 2.88 mm caudal and 15.6 mm rostral to the interaural plane, the horizontal planes situated between 16.5 and 22.5 mm from the brain's dorsal surface, and the sagittal planes between 0 and 6 mm from the medial wall. The coordinates correspond to area 29/31 in Paxinos et al. (2008, Wang et al., 2020).

We defined **VS** as lying within the coronal planes situated between 20.66 and 28.02 mm rostral to the interaural plane, the horizontal planes situated between 0 and 8.01 mm from the ventral surface of the striatum, and the sagittal planes between 0 and 8.69 mm from the medial wall. Note that our recording sites were targeted towards the nucleus accumbens core region of the VS. This dataset was used in Strait et al. (2015 and 2016).

We confirmed the recording location before each recording session using our Brainsight system with structural magnetic resonance images taken before the experiment. Neuroimaging was performed at the Rochester Center for Brain Imaging on a Siemens 3T MAGNETOM Trio Tim using 0.5 mm voxels or in the Center for Magnetic Resonance Research at UMN. We confirmed recording locations by listening for characteristic sounds of white and gray matter during recording, which in all cases matched the loci indicated by the Brainsight system.

**Electrophysiological techniques and processing:** Either single (FHC) or multi-contact electrodes (V-Probe, Plexon) were lowered using a microdrive (NAN Instruments) until waveforms between one and three neuron(s) were isolated. Individual action potentials were isolated on a Plexon system (Plexon, Dallas, TX) or Ripple Neuro (Salt Lake City, UT). Neurons were selected for study solely on the basis of the quality of isolation; we never preselected based on task-related response properties. All collected neurons for which we managed to obtain at least 300 trials were analyzed; no neurons that surpassed our isolation criteria were excluded from analysis.

**Eye-tracking and reward delivery:** Eye position was sampled at 1,000 Hz by an infrared eye-monitoring camera system (SR Research). Stimuli were controlled by a computer running Matlab (Mathworks) with Psychtoolbox and Eyelink Toolbox. Visual stimuli were colored rectangles on a computer monitor placed 57 cm from the animal and centered on its eyes. A standard solenoid valve controlled the duration of juice delivery. Solenoid calibration was performed daily.

**Risky choice task:** The task made use of vertical rectangles indicating reward amount and probability. We have shown in a variety of contexts that this method provides reliable communication of abstract concepts such as reward, probability, delay, and rule to monkeys (e.g. Azab et al., 2017 and 2018; Sleezer et al., 2016; Blanchard et al., 2014). The task presented two offers on each trial. A rectangle 300 pixels tall and 80 pixels wide represented each offer (11.35° of visual angle tall and 4.08° of visual angle wide). Two parameters defined gamble offers, *stakes* and *probability*. Each gamble rectangle was divided into two portions, one red and the other either gray, blue, or green. The size of the color portions signified the probability of winning a small (125 µl, gray), medium (165 µl, blue), or large reward (240 µl, green), respectively. We used a uniform distribution between 0 and 100% for probabilities. The size of the red portion indicated the probability of no reward. Offer types were selected at random with a 43.75% probability of blue (medium magnitude) gamble, a 43.75% probability of green (high magnitude) gambles, and a 12.5% probability of gray options (safe offers). All safe offers were excluded from the analyses described here, although we confirmed that the results are the same if these trials are included. Previous training history for these subjects included several saccade-based laboratory tasks, including a cognitive control task (Hayden et al., 2010), two stochastic choice tasks (Blanchard et al., 2014; Heilbronner and Hayden, 2016), a foraging task (Blanchard and Hayden, 2015), and a discounting task (Pearson et al., 2010).

On each trial, one offer appeared on the left side of the screen and the other appeared on the right. We randomized the sides of the first and second offer. Both offers appeared for 400 ms and were followed by a 600-ms blank period. After the offers were presented separately, a central fixation spot appeared, and the monkey fixated on it for 100 ms. Next, both offers appeared simultaneously and the animal indicated its choice by shifting gaze to its preferred offer and maintaining fixation on it for 200 ms. Failure to maintain gaze for 200 ms did not lead to the end of the trial but instead returned the monkey to a choice state; thus, monkeys were free to change their mind if they did so within 200 ms (although in our observations, they seldom did so). Following a successful 200-ms fixation, the gamble was resolved and the reward was delivered. We defined trials that took > 7 sec as inattentive trials and we did not include them in the analyses (this removed ~1% of trials). Outcomes that yielded rewards were accompanied by a visual cue: a white circle in the center of the chosen offer. All trials were followed by an 800-ms intertrial interval with a blank screen.

**Behavioral control for neural data indicates decisions from online valuation comparison:** The aim in this task design is to examine the process of valuation and decision-making in a sequential setting. In our design, the subjects are presented with the offers again to allow them the chance to reify their choice. To ensure that their choices are actually based on the online, sequential presentation of the offers in the main task, we examined previous behavioral data (in 7 sessions for each of 4 subjects) collected in a version of this task where the 2nd presentation does not occur, and they have to respond from memory. These results are also reported elsewhere (Yoo and Hayden, 2019). The result of these data was that subjects still chose the higher value offer approximately ~80% of the time. This result aligns with the findings with the main task data presented here, wherein (all) subjects chose the higher value approximately %82 of the time.

**Behavioral analysis, model estimation and subjective value:** The decision-variables underlying subject choice could arise from several possible estimates over probability, stakes and the estimated value. Our previous analysis and modeling of the present behavioral data indicate that the monkeys in this task make choices in line with a subjective valuation estimate of offers that reflects subjective attitudes towards the offer size (stakes) that indicate risk-seeking, and a warped probability estimation fit well by a prelec warping function. We remodeled the decisions here to derive subjective value estimates for using in estimation of neural encoding subspaces (see below) and establishing the connection of these subspaces to monkey choices. Here, we consider models where the monkey's subjective value (SV) for an offer follows the probability times the stakes.

 The model space we considered included 4 models. The models included those where the stakes were either assumed to be observed objectively or weighted as a power-law, and the probability was either observed objectively or transformed with a prelec function. For all models, choice probabilities were assumed to be generated by a softmax decision function over the relative subjective value. For example, during model fitting of a model with subjective utility (weighted stakes) and weighted probability, 3 model terms were fit. The terms for utility and probability, and the softmax temperature term. All models were fitted using the variational bayesian toolbox, and model comparisons performed using bayesian model selection over the model free energies (Daunizeau, Adam, & Rigoux, 2014).

**Full set of trial conditions:** The full simultaneous representation of the past and current offer can be described by eight distinct conditions, when we binarize offer value as before (see Methods). The values of the first and second offers are chosen independently – so, there are four possible combinations of the two offer values: both low, both high, and two combinations where one is high and the other is low (i.e., the left offer is high or the right offer is high). Then there are two possible orders: either the first offer is presented on the left (which means the second offer is presented on the right) or the first offer is presented on the right (and the second offer is presented on the left). This gives us eight conditions.

## Neural processing, data-selection, and statistical analysis

We calculated firing rates in 20-ms bins but we analyzed them in longer (400 ms) epochs. For estimating regression coefficients (see below), firing rates were all Z-scored for each neuron. In regression estimates, we only included risky trials. No smoothing was applied before any reported analysis.

**Single-neuron linear regression model:** Individual neuron selectivity was estimated using a linear regression model. Model coefficients were estimated using the z-scored, time-averaged firing rate in 4 different task windows. The windows were 400 ms blocks, with two separate blocks in the offer 1 window and 2 more blocks in the offer 2 window. The first blocks in a window were from 0 (offer on) to 400 ms after (offer off), and the second window was from 450 to 850 ms after the offer was presented. We refer to the latter window as the delay window. The estimated model had coefficients for subjective value, offer spatial position, and their interaction which can be used to estimate the nonlinear encoding terms. Model terms for offer side were effects coded [-1 1]; expected value was estimated with linear and b-spline terms (see **Figure S1** for more details on the splines), with the value term being min-max normalized to standardize across units and monkeys. The design matrix in total therefore had 5 variables to estimate plus the intercept. The model formula for a neuron (n) firing rate (FR) was estimated using all trials:

$$FR(n, trials) = \beta_0 + \beta_1 Value + \beta_2 Space + \beta(Interaction) + N(0, \sigma^2)$$

**Computing population encoding subspaces for space and value:** To measure separability between spatially distinct value subspaces, we used the computed regression coefficients for calculating the population encoding subspaces. Our goal was to compare the value subspaces for left and right offers both within offer epochs (e.g., offer 1) and across offer epochs. Because the coefficients ($\beta$) were derived from a linear model, we added up the coefficients to create a value vector for each distinct side and epoch.

Computation of each subspace involves setting the levels of X in the $\beta X$ of the regression equation, given an already set of fit $\beta$s per neuron (see *Single-Neuron linear regression model* above). Essentially, this process gives the model predicted firing rate spanning from the lowest range normalized value offer that includes the intercept offset (0) to the highest (1). The process of creating the value subspace vector for the left-side, for example, proceeded as follows: first, we must subtract two vectors, with the base vector centered on the different sides being compared (e.g., left and right), as they themselves have different intercepts. Given the linear model form for the design matrix, and assuming left == 1 and right == -1, the intercept vector for value ==0 and left offer subspace, the vector is:

$$Value\ high\ vector: [1]\beta_0 + [1]\beta_{Value} + [1]\beta_{Space} + [1]\beta_{Interaction}$$
$$-$$
$$Value/Side\ intercept\ vector: [1]\beta_0 + [0]\beta_{Value} + [1]\beta_{Space} + [0]\beta_{Interaction}$$

In total, there were six distinct task-functional comparisons, including comparisons of (1) left and right within offers 1 and 2, (2) left and right between offers 1 and 2, and (3) between the same-side across offers (i.e., left and left for offer 1 and 2, right and right for offer 1 and 2).

The testing of separability (or orthogonality) between subspaces requires establishing a null hypothesis of 1. This is because separable subspaces will be less correlated than a perfect correlation under the influence of noise. Specifically, because our main hypothesis was essentially a close to zero correlation between subspaces (i.e., $|r|$=0), we needed to estimate how a perfect correlation in the dataset, but confounded by noise, would be distributed ($|r|$=1). Therefore, we consider a subspace as separable or (semi-)orthogonal if the correlation between neuron weights is outside the confidence bound of this hypothetical perfect correlation distribution.

We addressed this problem by applying an already established approach (Kimmel et al., 2020). In brief, we fit the linear models described above on 1000 bootstrap resampled sets of trials. For each set, we computed the subspace correlation between left and right value representations. Then, to construct the noise ceiling estimate we computed the subspace correlation between the left value subspaces for the first and last 500 resamples (yielding 500 subspace correlation estimates) as well as the correlation between the right value subspaces from the first and last 500 resamples (yielding another 500 estimates). We then compared the 1000 subspace correlation estimates between the left and right value subspaces with the 1000 total estimates from the noise ceiling distribution using the bootstrap test.

Finally, we repeated this same procedure for the b-spline value representation models, using the alignment index (Kimmel et al., 2020) instead of the subspace correlation. These results are shown in **Figure S1**.

**Comparison between models with parallel and non-parallel subspaces:** To understand whether the single neuron basis of our semi-orthogonal subspaces could be explained equally well as noise, we performed a model comparison analysis for three classes of models: First, a regression model that fit only a noise term, and therefore explained the data only as random fluctuations. Second, models that had both side and value terms, but with no interaction between them. At the population level, this model would produce parallel subspaces. We fit these models with both linear and nonlinear representations of value (see **Figure S1** for detail). Third, models that have both side and value terms as well as an interaction term between them. If the interaction term is significant, then, at the population level, this model produces non-parallel subspaces.

For each model, we obtained a posterior predictive distribution, which is a distribution of samples from the model combined across all trials. We can use this posterior predictive distribution to check the fit of our model, and we find that our models provide good fit to the value response curves of the actual neurons (**Figures 2B** and **S3B**). Then, we perform a Bayesian model stacking analysis which assigns a weight to each model (Yao et al., 2017). In the stacking analysis, the posterior predictive distributions of all the models are combined to produce a combined model (i.e., a weighted sum of posterior predictive samples) that maximizes the leave-one-out cross-validation performance. We obtain similar results for more traditional measures of model goodness of fit, such as the widely applicable information criterion (Watanabe, 2013). However, we believe the stacking analysis provides useful intuition: It both gives insight into which model provides the best fit, as well as which combination of models provides the best fit.

**Neuron selectivity:** We determined the single neuron selectivity using a permutation ANOVA. For each neuron, we first fit an ANOVA using the same model terms as for the linear regression above. The only difference between the regression and ANOVA approach was that the ANOVA treated value as categorical and discretized over 7 equally distributed levels of value per monkey and per session. To control for chance detection of effects, we obtained a null distribution of ANOVA F-scores by re-running the ANOVA 1000 times per neuron using random permutations of the firing rate. The p-value was computed by for each effect and interaction as:

$$p = \frac{\#(Fperm > F_{main}) + 1}{\#(Fperm) + 1}$$

We used these p-values to determine whether a neuron was linearly selective for a single variable, linear mixed selectivity (both value and side terms significant), or nonlinearly selective (significant interaction term). This procedure was repeated across all 4 time windows, including both offer on epochs and the both offer delay windows. The proportion of each type of selectivity was determined using all monkeys within a brain region.

**Single-neuron basis of subspaces: bimodality:** To test whether the subspaces are primarily formed from gain modulating neurons, we note that the distribution of firing rate differences (across space) to value tuning should deviate from unimodal. We tested this by asking whether the distribution of differences in value tuning vectors for left and right offers was bimodal, using Hartigan's dip test of bimodality. Specifically, we formed distributions of tuning differences using the left and right value subspace coefficients for each neuron computed above. The distributional responses were computed for each offer (1 and 2) and time-window (offer on and delay) as a sensitivity index:

$$\text{Value-space firing rate (FR) sensitivity} = \frac{FR(n)_{left} - \mu(FR(n)_{left})}{\sigma(FR(n)_{left})} - \frac{FR(n)_{right} - \mu(FR(n)_{right})}{\sigma(FR(n)_{right})}$$

In the above, for example, $FR(n)_{left}$ is the value-left subspace vector coefficient for neuron $n$.

**Single-neuron basis of subspaces: subspace contribution of gain modulation versus complex nonlinear:** The bimodality test described above provides a global measure of whether the populations of neurons composing the subspaces are generally specialized to certain spatial locations. To further demarcate the contribution of gain versus complex nonlinear encodings to the subspaces, we use the property that gain modulated neurons will have the same preferred response to value, but with different amplitudes for different spatial positions, while complex nonlinear neurons will have a shifting preference to value. The main result we aimed for was determining the prevalence of nonlinear complex versus gain modulating neurons, and determining their % contribution of each type in forming the subspaces. To do this, we first determined which neurons in a region contributed to both left and right value subspaces for each epoch and time-window. Subspace contribution was determined by finding each neuron's percent variance in each subspace:

$$\text{Subspace contribution (\%)} = \frac{(FR(n)_{space}^2)^2}{\sum (FR(n)_{space}^4)}$$

The above neuron participation ratio is inherently related to the dimensionality of that subspace; effectively, each neurons % variance contributed can be viewed as an approximation to the eigenvalues of the subspace (Gao et al., 2017; Xie et al., 2022). We retained the top 95% of neurons, which was an average of 48 neurons in each subspace, across all brain regions. We then compared whether each neuron's preferred value was the same in left and right subspaces, within each of the analysis windows used for the regressions. Specifically, we performed a bootstrap hypothesis test of mean differences. The mean firing rates for testing were re-sampled 1000 times for each neuron, and the mean rate was computed over 7 equally distributed levels of value, separately for left and right offers. The peak rate for each resample was collected in a vector and subsequently randomized 1000 times (with resampling) to create a null distribution of mean differences in firing preference. The p-value was computed by counting the # of times the randomly permuted mean difference was larger than the empirically estimated.

## Binding by subspace orthogonalization

To motivate the problem we are studying here, we begin by considering a neural code that does not bind the distinct features of a single stimulus together. For example, a factorized representation of different stimulus features – even if the code for each individual feature is nonlinear – will fail to distinguish between different possible stimulus set bindings. In our setting, the neural population responds to both offer position and offer value. On a single trial the animal is shown a set of two stimuli, $X$, which each have a position and value. So, this set can be written as $X = \{[p_1, v_1], [p_2, v_2]\}$, where $p_1$ is the position of offer 1 and $v_1$ is its value. Then, if the average response of a neural population $\bar{r}$ is given by a function $r(X) = \sum_x^X f(x)$, we can write

$$\bar{r} \quad = r([p_1, v_1]) + f([p_2, v_2])$$

Now, if the function $f$ can be factorized into a sum of functions, $f_{pos}$ for offer position and $f_{val}$ for offer value, then we can rewrite the response as

$$\bar{r} \quad = f_{val}(v_1) + f_{pos}(p_1) + f_{val}(v_2) + f_{pos}(p_2)$$

This response contains all the original information about the features of the two offers – but it does not preserve their binding. In particular, while a maximum likelihood decoder would have a maximum at the correct stimulus set $X$, it would also have an equivalent maximum at the chimeric stimulus set $S = \{[p_1, v_2], [p_2, v_1]\}$. This is because the average population response to the two stimulus sets is exactly the same,

$$
\begin{aligned}
\Delta = \quad & r(X) - r(S) \\
= \quad & f_{val}(v_1) + f_{pos}(p_1) + f_{val}(v_2) + f_{pos}(p_2) \\
& -f_{val}(v_2) - f_{pos}(p_1) - f_{val}(v_1) - f_{pos}(p_2) \\
= \quad & 0
\end{aligned}
$$

Thus, in the case of a high-value offer on the left and a low-value offer on the right, an alternative and equally likely interpretation of the neural representation would be that there was a low-value offer on the left and a high-value offer on the right, which would lead to a suboptimal choice by the animal.

In practice, humans and other animals may make these kinds of binding errors, but they are thought to be infrequent (Bays et al., 2022). So, in many cases, this ambiguity must be successfully resolved. To understand how this happens, we consider the case when $f$ cannot be fully factorized and instead can be written as $f([p, v]) = f_{\text{pos}}(p) + f_{\text{val}}(v) + f_{\text{pos-val}}([p, v])$, where $f_{\text{pos-val}}$ is non-factorizable function of both side and value. We refer to this term as the conjunctive part of the response. Then, the difference in response for the correct and chimeric stimulus sets $\Delta$ from before can be written as,

$$\begin{aligned} \Delta \quad &= r(X) - r(S) \\ &= f_{\text{pos-val}}([v_1, p_1]) + f_{\text{pos-val}}([v_2, p_2]) - f_{\text{pos-val}}([v_2, p_1]) - f_{\text{pos-val}}([v_1, p_2]) \end{aligned}$$

Thus, so long as the squared sum of $\Delta$ across the neural population is larger than zero – and, in a noisy system, larger than the noise – then a decoder will be able to resolve the ambiguity between the correct and chimeric stimulus set and avoid misbinding errors. We derive an expression for this misbinding rate in a simplified case below. One of our key theoretical results is that the inclusion of this conjunctive part of the representation can simultaneously resolve the coding ambiguity without destroying the benefits of a factorized representation.

At the level of single neurons, the conjunctive part of the representation $f_{\text{pos-val}}$ manifests as neurons with nonlinear mixed selectivity for offer value and position. At the population level, the conjunctive part of the representation manifests as orthogonalization of the subspaces encoding the value of left and right offers.

**Subspace binding for continuous variables**

To make this point in a slightly different way, we consider a linear representation of two offer values. Here, we can write their encoding as

$$r(v) \quad = Av$$

where $v$ is a $2 \times 1$ vector of two offer values and $A$ is a $2 \times 2$ encoding matrix. We note that, in the noiseless case, we can recover the original vector $v$ exactly so long as $A$ has an inverse. When will $A$ have an inverse? Precisely when the two encoding vectors (the two columns of $A$) are not either perfectly parallel or perfectly anti-parallel. Formally,

$$A \quad = \begin{bmatrix} 1 & \cos(\theta) \\ 0 & \sin(\theta) \end{bmatrix}$$

where $\theta$ is the angle in radians between the encoding vectors, and $r = \cos(\theta)$ is the subspace correlation from the main text. The inverse of $A$ will not exist when $\sin(\theta) = 0$, which occurs when $\theta = 0, \pi$ and the subspace correlation $r = \pm 1$.

We can extend this simple analysis to the noisy case, with

$$r(v) \quad = Av + \eta$$

where $\eta \sim \mathbb{N}(0, \sigma)$. We are interested in the variance of the resulting estimate of $v$,

$$\begin{aligned} \text{Var}(\hat{v} - v) \quad &= \mathbb{E}_v[(A^{-1}Av + A^{-1}\eta - v)^2] \\ &= \mathbb{E}_v[(A^{-1}\eta)^2] \\ &= \frac{2\sigma^2}{\sin^2(\theta)} \end{aligned}$$

and so we can see that the variance of the estimate of $v$ is strictly decreasing with the absolute value of the subspace correlation.

# Linear-nonlinear code framework

To apply our mathematical framework to the experimental data, we consider a discretized discretized offer value along with offer position – value is discretizd as for our decoding analysis (see below). In these data, we found that decoding a binary value yielded much better performance than decoding continuous value with various methods. The analytic approach we take below is also simplified for discrete value. However, the subspace binding hypothesis does not depend on this discreteness, as discussed above. Once we discretize value, we have $K = 2$ latent variables that each take on $n = 2$ different values. However, the theory we develop applies to any choice of $K$ and $n$, and so can be applied well beyond this experimental setting.

We model the neural responses of $N$ neurons as,

$$r(x) = Lx_z + Mf_N(x) + \epsilon$$

where $r(x)$ is an $N \times 1$ vector of the z-scored activity of $N$ neurons, $L$ is an $N \times K$ linear transform of the z-scored stimulus vector $x_z$ and has columns $L = [d_{LV}L_1 \; d_{LA}L_2]$, $M$ is an $N \times n^K$ linear transform of $f_N(x)$, which is a nonlinear transform of the stimulus vector $x$ (defined in detail below). Finally, $\epsilon \sim \mathcal{N}(0, \sigma^2)$. A note: We z-score the stimulus vector $x$ (defined in detail below), so that the linear distance in the representation produced after the linear transform is controlled fully by the linear transform $L$, and does not depend on the specific choice of coding for $x$.

In the experiment, there were two features that each took on two values. So, $x$ is a vector with two elements that are each either 1 or 2. The first element corresponds to low- (1) or high-value (2); the second element corresponds to the two values of offer position. The possible $x$ are:

$$x_{11} = [1 \; 1]^T$$
$$x_{12} = [1 \; 2]^T$$
$$x_{21} = [2 \; 1]^T$$
$$x_{22} = [2 \; 2]^T$$

The nonlinear function we consider is the conjunctive identity function used in (Rigotti et al., 2013; Johnston et al., 2020), where

$$f_N(x)_{ij} = [x_i = i][x_j = j]$$

and

$$f_N(x) = [f_N(x)_{11} \; f_N(x)_{12} \; f_N(x)_{21} \; f_N(x)_{22}]^T$$

For the linear transform $L$, which has $K = 2$ columns, the length of the first column is $d_{LV}$, which will be the average distance between two stimulus representations that differ only in their value (where $M = 0$). We will refer to the length of the second column as $d_{LA}$, the average distance between two stimuli that differ only in offer position (also where $M = 0$). All of the columns of $M$ will have the same length $m$, which will mean that the distance between the nonlinear components of two stimulus representations will be $d_N = \sqrt{2}m$. We assume that the length of the nonlinear perturbation for each stimulus is the same; while this is unlikely to be precisely true, it simplifies our analysis and still gives good results when comparing to the experimental data. We also assume that the columns of both $L$ and $M$ are orthogonal, both within each matrix and between the two matrices (though the analytic results are similar without the between matrix orthogonality). For large $N$, this will tend to be true for random vectors.

Finally, a code in this framework is described by a stimulus set ($K$ and $n$) and four parameters: $d_{LV}$ (representing offer value), $d_{LA}$ (representing offer position), $d_N$ (representing the nonlinear part of the code), and $\sigma$ (representing the standard deviation of the noise). In our analytic theory, we show that the binding error rate depends on $d_N$ and $\sigma$. That is, low binding error rates can be achieved so long as $d_N$ is sufficiently large, and does not depend on the linear part of the code ($d_{LV}$ or $d_{LA}$). We also show that the generalization error rate depends only on $d_{LV}$, $d_N$, and $\sigma$ – that is, it does not depend on $d_{LA}$. Thus, we will focus on estimating $d_{LV}$, $d_N$, and $\sigma$ from the experimental data.

**Linear-nonlinear code with constrained power**

We will consider linear-nonlinear codes that have different allocations of power between their linear and nonlinear parts. To do this, we want to keep the total power $P$ constant and we vary the linear power $P_L$ and nonlinear power $P_{NL}$. In particular, $P = P_L + P_{NL}$. To understand the implications of a particular combination of $P_L$ and $P_{NL}$, we need to understand the relationship between a particular $P$ and the corresponding distance.

**Linear distance:** We derive the distance between adjacent stimuli in the linear code for a particular number of features $K$ and number of values $n$ that each feature takes on as well as the linear power of the code ($P_L$),

$$d_L = \sqrt{\frac{12 P_L}{K(n^2 - 1)}}$$

We approach this by computing the variance (i.e., the linear power $P_L$) of a uniformly sampled $K$-dimensional lattice with $n$ points spaced at distance $d_L$ along each dimension. Then, we invert the expression for the variance to find an expression for the distance between the points. First, we write the variance $P_L$ as

$$n^K P_L = \sum_{i=0}^{n-1} \left[ \left( i - \frac{n-1}{2} \right)^2 d_L^2 + \sum_{j=0}^{n-1} \left[ \left( j - \frac{n-1}{2} \right)^2 d_L^2 + \ldots \right] \right]$$

$$= \sum_{i}^{n-1} \sum_{j}^{n-1} \ldots \sum_{k}^{n-1} \left( i - \frac{n-1}{2} \right)^2 d_L^2 + \left( j - \frac{n-1}{2} \right)^2 d_L^2 + \ldots + \left( k - \frac{n-1}{2} \right)^2 d_L^2$$

$$= K n^{K-1} \sum_{i}^{n-1} \left( i - \frac{n-1}{2} \right)^2 d_L^2$$

$$= K n^{K-1} d_L^2 \sum_{i}^{n-1} i^2 - (n-1) \sum_{i}^{n-1} i + n \frac{n-1}{2}^2$$

and we can rewrite this with known expressions for the sum of integers and sum of squared integers up to a particular value,

$$n^K P_L = K n^{K-1} d_L^2 \left[ \frac{(n-1)n(2n-1)}{6} - \frac{n(n-1)^2}{2} + \frac{n(n-1)^2}{4} \right]$$

$$= K n^K d_L^2 \left[ \frac{(n-1)(2n-1)}{6} - \frac{(n-1)^2}{4} \right]$$

$$= K n^K d_L^2 \left[ \frac{2n^2 - 3n + 1}{6} - \frac{n^2 - 2n + 1}{4} \right]$$

$$= K n^K d_L^2 \left[ \frac{4n^2 - 6n + 2}{12} - \frac{3n^2 - 6n + 3}{12} \right]$$

$$n^K P_L = K n^K d_L^2 \frac{n^2 - 1}{12}$$

$$P_L = K d_L^2 \frac{n^2 - 1}{12}$$

Now, we rewrite in terms of $d_L$,

$$d_L = \sqrt{\frac{12 P_L}{K(n^2 - 1)}}$$

which is the expression given above.

Following from the lattice structure, stimuli at a diagonal point on the lattice have distance $\sqrt{2} d_L$.

**Linear neighbors derivation:** Second, we find the average number of neighbors that a particular stimulus has at both this nearest distance $N_{LA}$ and nearest diagonal distance $N_{LD}$. This is a counting problem. We observe that, in the lattice, there are two edge values for each feature and $n - 2$ non-edge values. Thus,

$$N_{LA} = \frac{1}{n^K} \sum_{c=0}^{K} (2K - c) \binom{K}{c} (n-2)^{K-c} 2^c$$

and

$$N_{LD} = \frac{1}{n^K} \sum_{c=0}^{K} \left( 4 \binom{K-c}{2} + 2(K-c)c + \binom{C}{2} \right) \binom{K}{c} (n-2)^{K-c} 2^c$$

**Nonlinear distance:** The nonlinear distance is

$$d_N = \sqrt{2 P_N}$$

and has been treated in more detail previously (Johnston et al., 2020).

**Nonlinear neighbors derivation:** Because each nonlinear representation is along a vector that is orthogonal to all other nonlinear representations, from a particular stimulus all other representations are at minimum distance. So,

$$N_N L = n^K - 1$$

**Total code distance:** The total code distance for orthogonal linear and nonlinear code parts is

$$d_C = \sqrt{d_L^2 + d_{NL}^2}$$

If the linear and nonlinear code parts are not constrained to be orthogonal, then the total code distance is a random variable with the following form,

$$d_C = \sqrt{d_L^2 + d_{NL}^2 + 2d_{NL}d_L\eta}$$

where $\eta \sim \mathcal{N}(0, 1/N)$ due to the fact that the dot product of two unit vectors are normally distributed with variance inverse to their length (i.e., the distribution of $\eta$). The distance is similary defined for the next nearest stimuli,

$$d_{C+1} = \sqrt{2d_L^2 + d_{NL}^2 + \sqrt{8}d_{NL}d_L\eta}$$

**Total code neighbors:** To combine the code neighbors, it is enough to simply take the minimum between the linear and nonlinear parts, which will always be equal to $N_{LA}$ (or $N_{LD}$ for next-nearest). So,

$$N_C = N_{LA}$$
$$N_{C+1} = N_{LD}$$

**Relating subspace correlation to linear and nonlinear distance**

The linear-nonlinear code framework developed here provides a different way to define the subspace correlation measured used in the rest of the paper. We compute subspace correlation directly from the linear ($d_{LV}$) and nonlinear distances ($d_N$). In particular, we can take the cosine similarity between $v_l$ and $v_r$, as defined in the main text,

$$\rho = \frac{v_1 \cdot v_2}{|v_1|_2 |v_2|_2}$$
$$= \frac{(L_1 + M_1 - M_2) \cdot (L_1 + M_3 - M_4)}{d_{LV}^2 + d_N^2}$$
$$= \frac{d_{LV}^2}{d_{LV}^2 + d_N^2}$$

where $\rho$ is the subspace correlation (and see **Figure 3A** for a schematic).

**The overall error rate of a linear-nonlinear code**

Using all of the expressions developed in the previous section, we can write an approximation for the likelihood that a linear-nonlinear code makes an error, for a given stimulus set, defined by $K$ and $n$, and given linear and nonlinear powers $P_L$ and $P_{NL}$.

Here, we define an error as the most likely stimulus (set) under a maximum likelihood decoder $\hat{x}$ not being the original stimulus (set) $x$. We take the a nearest neighbor union bound approach and develop the following expression for the error rate,

$$P(\text{error}) = N_C Q\left(-\frac{d_C}{2\sigma}\right) + N_{C+1} Q\left(-\frac{d_{C+1}}{2\sigma}\right)$$

where $Q(.)$ is the standard Gaussian cdf. Thus, we can see that the error rate depends most strongly on the distances. From our distance definitions, we know that, if $K > 2$ or $n > 2$, increasing nonlinear power is the most efficient way to increase distance. As a consequence, to drive the error rate down, it is best to put all code power toward the nonlinear part.

For multiple stimuli, the code error rate can be written as,

$$P(\text{error}) \quad = SN_C Q\left(-\frac{d_C}{2\sigma}\right) + SN_{C+1} Q\left(-\frac{d_{C+1}}{2\sigma}\right) + P(\text{binding error})$$

where $S$ is the number of stimuli. We will develop an approximation for $P(\text{binding error})$ in more detail in the next section.

**Derivation of the binding error rate**

We begin by considering a purely linear code for multiple stimuli $X$, where

$$r_L(X) = \sum_{x \in X} L\, x_z + \epsilon$$

where $x_z$ is the z-scored features of $x$ and $L$ is an $N \times K$ linear transform – as above. With a purely linear code and stimuli that are defined by two features $K = 2$ that each take on two values $n = 2$, there are two stimulus pairs that give rise to exactly the same average response. Those pairs are

$$X = \{x_{11}, x_{22}\}$$
$$S = \{x_{12}, x_{21}\}$$

which is to say,

$$\bar{r}_L(X) \quad = \bar{r}_L(S)$$

Due to this property, even a decoder that is optimal for our current setting (the maximum likelihood decoder), will not be able to discriminate between these two options at an error rate different from chance. If $X$ is presented, then we refer to trials in which $S$ is decoded from the activity by a maximum likelihood decoder as a misbinding error. How can we modify the code to make fewer misbinding errors? Here, we show that reintroducing the nonlinear part of the code can effectively drive down the probability of binding errors.

Now, we return to the full code, given above. We want to show how increasing the nonlinear distance $d_N$ decreases the probability of misbinding errors. First, we derive the distance between the average representation of $X$ and $S$ in the full code (that is, with non-zero nonlinear distance), where,

$$r(X) = \sum_{x \in X} L\, x_z + M f_N(x) + \epsilon$$

So, for $X$ and $S$ defined above,

$$
\begin{aligned}
d_S \quad &= |\bar{r}(X) - \bar{r}(S)|_2 \\
&= \left| \sum_{x \in X} L\, x_z + M f_N(x) - \sum_{s \in S} L\, s_z - M f_N(s) \right|_2 \\
&= \left| \sum_{x \in X} M\, f_N(x) - \sum_{s \in S} M\, f_N(s) \right|_2 \\
&= |M_1 + M_2 - M_3 - M_4|_2 \\
&= \sqrt{2} d_N
\end{aligned}
$$

We notice that this does not depend on the particular stimulus sets $X$ and $S$ anymore. Indeed, due to our assumption that the nonlinear distances are constant across stimuli, this is the difference between any two sets of stimuli that are linearly confusable.

Now, knowing the distance between the correct stimulus set $X$ and a misbound stimulus set $S$, $d_S$, we can use the following expression for the rate of binding errors via a union bound approximation where $S$ remains a particular stimulus set that is linearly confusable with $X$, $C_X$ is the set of all stimulus sets that are linearly confusable with $X$, and $\hat{X}$ is the stimulus set inferred by a maximum likelihood decoder:

$$
\begin{aligned}
P\left(\hat{X} \in C_X \mid r(X)\right) \quad &= P\left(\underset{S \in C_X}{\cup} \hat{X} = S \mid r(X)\right) \\
&\leq \sum_{S \in C_X} P\left(\hat{X} = S \mid r(X)\right) \\
&= \sum_{S \in C_X} P\left(d\left(\bar{r}(S), r(X)\right) < d\left(\bar{r}(X), r(X)\right)\right) \\
&\approx \sum_{S \in C_X} Q\left(-\frac{d_S}{2\sigma}\right)
\end{aligned}
$$

where $d(.,.)$ takes the Euclidean distance between its two arguments and $Q(.)$ is the standard Gaussian cdf. Then, we average over $X$,

$$
P(\text{binding error}) \quad \approx N_S \, Q\left(-\frac{d_S}{2\sigma}\right)
$$

where $N_S$ is the average size of $C_X$ across all $X$. In our case, for two stimuli with two features that each take on two values, $N_S = \frac{1}{4}$. In general, we can write the number of chimeric stimulus sets as

$$
N_S \quad = \frac{1}{2}\binom{S}{2}\sum_{i=0}^{K-1}\binom{K}{i}\frac{1}{n}^i\left(1-\frac{1}{n}\right)^{K-i}\sum_{j=1}^{K-i}\binom{K-i}{i}
$$

**Derivation of the generalization error rate**

Next, we want to develop a prediction for the generalization error rate that depends on our four parameters $d_{LV}$, $d_N$, and $\sigma$ as well as on the standard error of the mean $\epsilon$. In particular, we can compare the generalization error rate predicted by our theory and developed here from the generalization error rate of decoders trained on the neural data (discussed more below). This provides a crucial validation test for our theory, and the close correspondence observed between the theory and the data indicate that our formalization captures the relevant aspects of the geometry of the neural representations.

Here, we will develop the approximation assuming that the linear and nonlinear parts of the code are orthogonal to each other. However, the results are similar if the linear and nonlinear parts are randomly chosen with respect to each other.

As before, we have four stimuli of interest $x_{ij}$ for $i, j \in \{1,2\}$. We can write the representation corresponding to each of them in terms of the linear and nonlinear code components where $M_i$ denotes the columns of $M$ and $L_i$ denotes the columns of $L$, such that,

$$
\begin{aligned}
r(x_{11}) \quad &= 0 \\
r(x_{21}) \quad &= d_{LV}L_1 + d_{N\epsilon}M_{12} \\
r(x_{12}) \quad &= d_{LA}L_2 + d_{N\epsilon}M_{13} \\
r(x_{22}) \quad &= d_{LV}L_1 + d_{LA}L_2 + d_{N\epsilon}M_{14}
\end{aligned}
$$

where $d_{LA}$ is the linear distance associated with the variable that is being generalized across (i.e., offer side or time), $d_{N\epsilon}$ is the combined distance from both the nonlinear part of the code and the SEM $d_{N\epsilon} = \sqrt{d_N^2 + \epsilon^2}$, and we have defined

$$M_{ij} = \frac{M_i - M_j}{\sqrt{2}}$$

as well as for convenience. The standard error distance $\epsilon$ appears here because some of the perturbations to the underlying linear structure are reliable and form the nonlinear distance, while others are unreliable and emerge due to the noisy estimation of each centroid. The generalization performance of the classifier is reduced by both – while, for instance, the traditional cross-validated performance of the classifier would be reduced only by the standard error of the mean.

Next, we consider a linear prototype decoder – that is, a decoder that returns a binary classification of a new point by decoding whether it is closer to the class centroid for the first or second class. In this case, the two classes are separated by a linear hyperplane, which is orthogonal to the vector that connects the two class. This method of learning the separating hyperplane does not maximize the margin between the two classes, as done by support vector machines. Here, we compute the generalization error rate for a putative prototype decoder. Later, we show that this generalization performance is a good match for the generalization performance of a support vector machine decoder.

As mentioned above, the prototype decoder uses the vector connecting the two class centroids to decode new points. In our framework, we can write an expression for this vector given the two stimuli $x_{11}$ and $x_{21}$,

$$v_D = \frac{1}{c}\big(r(x_{21}) - r(x_{11})\big)$$
$$= \frac{1}{c}(d_L L_1 + d_{N\epsilon} M_{12})$$

where $c$ normalizes $v_D$ to be a unit vector and has the form

$$c = \sqrt{d_L^2 + d_{N\epsilon}^2}$$

Because we assume that the noise magnitude is the same for each stimulus, the decoding boundary is $\frac{c}{2}$. In particular, to decode an unseen data point $y$, we would evaluate,

$$o = \text{sgn}\left[v_D \cdot y - \frac{c}{2}\right]$$

where sgn takes the sign of its argument. If $o = -1$, then we would classify $y$ as, for instance, low value; if $o = 1$, then we would classify $y$ as high value.

So, to evaluate the generalization performance of this decoder on held out stimulus conditions, we can apply this same logic to the left out stimuli $x_{12}$ and $x_{22}$. In particular, we want to project the representations of these held out stimuli onto the decoder vector $v_D$ derived above and then compare the position of the representations along that vector to the decoding threshold $\frac{c}{2}$. So, for $x_{12}$,

$$\begin{aligned}
d_{12} &= v_D \cdot r(x_{12}) - \frac{c}{2} \\
&= \frac{1}{c}(d_{LV}L_1 + d_{N\epsilon}M_{12}) \cdot (d_{LA}L_2 + d_{N\epsilon}M_{13}) - \frac{c}{2} \\
&= \frac{1}{c2}d_{N\epsilon}^2 - \frac{c}{2} \\
&= \frac{1}{c2}d_{N\epsilon}^2 - \frac{c^2}{c2} \\
&= \frac{d_{N\epsilon}^2}{c2} - \frac{d_{LV}^2 + d_{N\epsilon}^2}{c2} \\
&= -\frac{\frac{1}{2}d_{LV}^2}{\sqrt{d_{LV}^2 + d_{N\epsilon}^2}}
\end{aligned}$$

and, for $x_{22}$,

$$\begin{aligned}
d_{22} &= v_D \cdot r(x_{22}) - \frac{c}{2} \\
&= \frac{1}{c}(d_{LV}L_1 + d_{N\epsilon}M_{12}) \cdot (d_{LA}L_2 + d_{LV}L_1 + d_{N\epsilon}M_{14}) - \frac{c}{2} \\
&= \frac{1}{c}\left(d_{LV}^2 + \frac{1}{2}d_{N\epsilon}^2\right) - \frac{c}{2} \\
&= \frac{\frac{1}{2}d_{LV}^2}{\sqrt{d_{LV}^2 + d_{N\epsilon}^2}}
\end{aligned}$$

Now, using these two distances and the noise magnitude $\sigma$, we can predict how well a decoder trained to discriminate $x_{11}$ from $x_{21}$ will generalize to discriminate $x_{12}$ from $x_{22}$. This is the cross-category generalization performance from (Bernardi et al., 2020) and that is discussed in the main text. In particular, the

$$\begin{aligned}
P(\text{CCGP error}) &\approx \frac{1}{2}Q\left(\frac{d_{12}}{\sigma}\right) + \frac{1}{2}Q\left(-\frac{d_{22}}{\sigma}\right) \\
&= Q\left(-\frac{1}{\sigma}\frac{\frac{1}{2}d_{LV}^2}{\sqrt{d_{LV}^2 + d_{N\epsilon}^2}}\right)
\end{aligned}$$

where $Q$ is the cumulative distribution function of a standard normal distribution, as before.

## Linear-nonlinear data decomposition

We want to estimate the three parameters of our code model, $d_{LV}$ (the linear code distance for value), $d_N$ (the nonlinear code distance), and $\sigma$ (the noise standard deviation), from the data.

To do this, we use a cross-validated distance measure, often referred to as the crossnobis distance (though we do not incorporate an estimate of the noise structure into our use of the measure, so it is not a version of the Mahalanobis distance as the name crossnobis suggests), to estimate the unbiased euclidean distance between every pair of our four stimulus conditions. We used the routines provided by the python RSA toolbox (Nili et al., 2014). This yields a representational distance matrix, which can be decomposed according to our model framework.

In particular, we can frame this as a constrained least squares optimization problem. We want to find the solution to

$$d_{\text{empirical}} = A \, d_{\text{est}}$$

where $d_{\text{empirical}}$ is a flattened version of the crossnobis distance matrix estimated above (now a $6 \times 1$ vector), $A$ is a $6 \times 3$ design matrix where each row gives the integer multiples of the linear and nonlinear distances between the two corresponding conditions (for instance, conditions that have the same value, but different positions would have a row vector $[0 \; 1 \; 2]$ in the matrix while conditions with different values and different positions would have $[1 \; 1 \; 2]$), and $d_{\text{est}} = [d_{LV}^2 \; d_{LA}^2 \; d_N^2]^T$. Then, we find the least squares solution to the above equation with $d_{\text{est}}$ constrained to be non-negative. See Figure S3 for this procedure applied to synthetic data, with a similar number of units and noise level as in our recordings – it accurately recovers both the linear and nonlinear distances.

Next, we estimate the magnitude of the noise in the neural representations. Here, we choose to estimate it along a single dimension that is particularly relevant for our analyses. For our generalization analysis, $x_{11}$ and $x_{21}$ are defined as the training set while $x_{12}$ and $x_{22}$ are defined as the testing set. That is, the decoder will be trained to discriminate between $r(x_{11})$ and $r(x_{21})$ and then tested on its ability to discriminate between $r(x_{12})$ and $r(x_{22})$. Due to this, noise specifically along this learned decoding dimension is most relevant to generalization performance, and we will estimate specifically the magnitude of this noise from our data. In particular,

$$\sigma^2 = \mathbb{E}_{x_{ij}} \left[ \frac{v_1}{|v_1|_2} \cdot \left( r(x_{ij}) - \bar{r}(x_{ij}) \right) \right]^2$$

where the expectation is taken across both all trials from a particular condition and all conditions defined by $i$ and $j$.

Finally, we also compute the standard error of the mean of each of our estimates, collapsed into a standard error distance $\epsilon$, which will affect generalization performance, as described below.

## Decoding analyses

### Binarizing value

We discretize value into high and low by splitting it according to the subjective value transformation computed for each session and excluding the middle 30 percentile.

### Preprocessing the neural data

Before decoding, we preprocess the data by z-scoring and then applying a PCA that retains enough dimensions to capture 99 % of the variance. Both the z-score and PCA transforms are fit on the training set only. All decoding analyses are done on data from three non-overlapping 300 ms bins that begin 100 ms after offer onset. The activity from each neuron in each time bin are treated as separate features. All of the bins from a single trial are in either the training or testing set, they are not split across both.

For decoding, we consider value presented in two distinct conditions. We begin by constructing pseudopopulations for high- and low-value in each condition (e.g., high- and low-value on the left and high- and low-value on the right). The pseudopopulation consists of all neurons from a particular brain region with at least 160 trials for each of the four conditions for the broad data splits (e.g., splitting into left and right presentations with high or low value, combining across offer 1 and offer 2) and at least 80 trials for the narrower data splits (e.g., splitting into high and low value for left offer 1 and right offer 2).

**Decoding**

Then, we train a support vector machine decoder with a linear kernel to discriminate high- from low-value in one condition (e.g., decoding value from only presentation on the left) and test that decoder both on held-out trials from that condition (10 % of trials are held out) and all trials from the other condition (e.g., high- and low-value on the right). The performance of the decoder on the held out trials from the training condition is the standard decoding performance and the performance of the decoder on the trials from the second condition is the cross-condition generalization performance.

All decoding analyses are implemented in scikit-learn (Pedregosa et al., 2011).

In the main text, we compare the generalization performance of these SVM decoders to the generalization performance that is predicted by our analysis of the linear-nonlinear code model, given above. We believe that the close correspondence observed between the empirical and predicted generalization performance indicates that our linear-nonlinear code formalization captures relevant aspects of the neural representation geometry.

# REFERENCES

Aoi, M. C., Mante, V., & Pillow, J. W. (2020). Prefrontal cortex exhibits multidimensional dynamic encoding during decision-making. Nature neuroscience, 23(11), 1410-1420.

Babadi, B., & Sompolinsky, H. (2014). Sparseness and expansion in sensory representations. Neuron, 83(5), 1213-1226.

Barak, O., Rigotti, M., & Fusi, S. (2013). The sparseness of mixed selectivity neurons controls the generalization–discrimination trade-off. Journal of Neuroscience, 33(9), 3844-3856.

Bays, P., Schneegans, S., Ma, W. J., & Brady, T. (2022). Representation and computation in working memory. psyRxiv preprint: https://doi.org/10.31234/osf.io/kubr9

Bernardi, S., Benna, M. K., Rigotti, M., Munuera, J., Fusi, S., & Salzman, C. D. (2020). The geometry of abstraction in the hippocampus and prefrontal cortex. Cell, 183(4), 954-967.

Blanchard, T. C., Piantadosi, S. T., & Hayden, B. Y. (2018). Robust mixture modeling reveals category-free selectivity in reward region neuronal ensembles. Journal of neurophysiology, 119(4), 1305-1318.

Blanchard, T. C., Hayden, B. Y., & Bromberg-Martin, E. S. (2015). Orbitofrontal cortex uses distinct codes for different choice attributes in decisions motivated by curiosity. Neuron, 85(3), 602-614.

Blanchard, T. C., Wolfe, L. S., Vlaev, I., Winston, J. S., & Hayden, B. Y. (2014). Biases in preferences for sequences of outcomes in monkeys. Cognition, 130(3), 289-299.

Botvinick, M., & Watanabe, T. (2007). From numerosity to ordinal rank: a gain-field model of serial order representation in cortical working memory. Journal of Neuroscience, 27(32), 8636-8642.

Boyle, L., Posani, L., Irfan, S., Siegelbaum, S. A., & Fusi, S. (2022). The geometry of hippocampal CA2 representations enables abstract coding of social familiarity and identity. *bioRxiv*.

Cai, X., & Padoa-Schioppa, C. (2014). Contributions of orbitofrontal and lateral prefrontal cortices to economic choice and the good-to-action transformation. Neuron, 81(5), 1140-1151.

Chung, S., & Abbott, L. F. (2021). Neural population geometry: An approach for understanding biological and artificial networks. Current opinion in neurobiology, 70, 137-144.

Cueva, C. J., Saez, A., Marcos, E., Genovesio, A., Jazayeri, M., Romo, R., ... & Fusi, S. (2020). Low-dimensional dynamics for working memory and time encoding. Proceedings of the National

Academy of Sciences, 117(37), 23021-23032.

Daunizeau J, Adam V, Rigoux L (2014) VBA: A Probabilistic Treatment of Nonlinear Models for Neurobiological and Behavioural Data. PLOS Computational Biology 10(1): e1003441

Dean, H. L., & Platt, M. L. (2006). Allocentric spatial referencing of neuronal activity in macaque posterior cingulate cortex. Journal of Neuroscience, 26(4), 1117-1127.

Dosher, B., & Lu, Z. L. (2017). Visual perceptual learning and models. Annual review of vision science, 3, 343-363.

Ebitz, R. B., & Hayden, B. Y. (2021). The population doctrine in cognitive neuroscience. Neuron, 109(19), 3055-3068.

Eliasmith, C., Stewart, T. C., Choo, X., Bekolay, T., DeWolf, T., Tang, Y., & Rasmussen, D. (2012). A large-scale model of the functioning brain. Science, 338(6111), 1202-1205.

Elsayed, G. F., Lara, A. H., Kaufman, M. T., Churchland, M. M., & Cunningham, J. P. (2016). Reorganization between preparatory and movement population responses in motor cortex. Nature communications, 7(1), 1-15.

Farashahi, S., Azab, H., Hayden, B., & Soltani, A. (2018). On the flexibility of basic risk attitudes in monkeys. Journal of Neuroscience, 38(18), 4383-4398.

Farashahi, S., Donahue, C. H., Hayden, B. Y., Lee, D., & Soltani, A. (2019). Flexible combination of reward information across primates. Nature human behaviour, 3(11), 1215-1224.

Feierstein, C. E., Quirk, M. C., Uchida, N., Sosulski, D. L., & Mainen, Z. F. (2006). Representation of spatial goals in rat orbitofrontal cortex. Neuron, 51(4), 495-507.

Fine, J. M., & Hayden, B. Y. (2022). The whole prefrontal cortex is premotor cortex. Philosophical Transactions of the Royal Society B, 377(1844), 20200524.

Flesch, T., Juechems, K., Dumbalska, T., Saxe, A., & Summerfield, C. (2022). Orthogonal representations for robust context-dependent task performance in brains and neural networks. Neuron.

Fusi, S., Miller, E. K., & Rigotti, M. (2016). Why neurons mix: high dimensionality for higher cognition. Current opinion in neurobiology, 37, 66-74.

Gallego, J. A., Perich, M. G., Miller, L. E., & Solla, S. A. (2017). Neural manifolds for the control of movement. Neuron, 94(5), 978-984.

Gore, F., Hernandez, M., Ramakrishnan, C., Crow, A. K., Malenka, R. C., & Deisseroth, K.

(2023). Orbitofrontal cortex control of striatum leads economic decision-making. *Nature Neuroscience*, 1-9.

Greff, K., Van Steenkiste, S., & Schmidhuber, J. (2020). On the binding problem in artificial neural networks. arXiv preprint arXiv:2012.05208.

Hare, T. A., Schultz, W., Camerer, C. F., O'Doherty, J. P., & Rangel, A. (2011). Transformation of stimulus value signals into motor commands during simple choice. Proceedings of the National Academy of Sciences, 108(44), 18120-18125.

Hayden, B. Y., & Moreno-Bote, R. (2018). A neuronal theory of sequential economic choice. Brain and Neuroscience Advances, 2, 2398212818766675.

Hayden, B. Y. (2019). Why has evolution not selected for perfect self-control?. Philosophical Transactions of the Royal Society B, 374(1766), 20180139.

Hayden, B. Y., & Niv, Y. (2021). The case against economic values in the orbitofrontal cortex (or anywhere else in the brain). Behavioral Neuroscience, 135(2), 192.

Hocker, D. L., Brody, C. D., Savin, C., & Constantinople, C. M. (2021). Subpopulations of neurons in lOFC encode previous and current rewards at time of choice. Elife, 10, e70129.

Jazayeri, M., & Ostojic, S. (2021). Interpreting neural computations by examining intrinsic and embedding dimensionality of neural activity. Current opinion in neurobiology, 70, 113-120.

Johnston, W. J., & Freedman, D. J. (2023). Redundant representations are required to disambiguate simultaneously presented complex stimuli. PLOS Computational Biology, 19(8), e1011327.

Johnston, W. J., & Fusi, S. (2023). Abstract representations emerge naturally in neural networks trained to perform multiple tasks. Nature Communications, 14(1), 1040.

Johnston, W. J., Palmer, S. E., & Freedman, D. J. (2020). Nonlinear mixed selectivity supports reliable neural computation. PLOS computational biology, 16(2), e1007544.

Jonikaitis, D., & Zhu, S. (2023). Action space restructures visual working memory in prefrontal cortex. bioRxiv, 2023-08.

Kable, J. W., & Glimcher, P. W. (2007). The neural correlates of subjective value during intertemporal choice. Nature neuroscience, 10(12), 1625-1633.

Kable, J. W., & Glimcher, P. W. (2009). The neurobiology of decision: consensus and controversy. Neuron, 63(6), 733-745.

Kaufman, M. T., Benna, M. K., Rigotti, M., Stefanini, F., Fusi, S., & Churchland, A. K. (2022). The implications of categorical and category-free mixed selectivity on representational geometries. Current Opinion in Neurobiology, 77, 102644.

Kimmel, D. L., Elsayed, G. F., Cunningham, J. P., & Newsome, W. T. (2020). Value and choice as separable and stable representations in orbitofrontal cortex. Nature communications, 11(1), 1-19.

Knudsen, E. B., & Wallis, J. D. (2022). Taking stock of value in the orbitofrontal cortex. Nature reviews. Neuroscience, 10.1038/s41583-022-00589-2.

Krajbich, I., Armel, C., & Rangel, A. (2010). Visual fixations and the computation and comparison of value in simple choice. Nature neuroscience, 13(10), 1292-1298.

Libby, A., & Buschman, T. J. (2021). Rotational dynamics reduce interference between sensory and memory representations. Nature neuroscience, 24(5), 715-726.

Litwin-Kumar A, Harris KD, Axel R, Sompolinsky H, Abbott LF. Optimal Degrees of Synaptic Connectivity. Neuron. 2017;0(0):1153–1164.e7.

Luk, C. H., & Wallis, J. D. (2013). Choice coding in frontal cortex during stimulus-guided or action-guided decision-making. Journal of Neuroscience, 33(5), 1864-1871.

Maggi, S., & Humphries, M. D. (2022). Activity subspaces in medial prefrontal cortex distinguish states of the world. Journal of Neuroscience, 42(20), 4131-4146.

Mante, V., Sussillo, D., Shenoy, K. V., & Newsome, W. T. (2013). Context-dependent computation by recurrent dynamics in prefrontal cortex. nature, 503(7474), 78-84.

Matthey, L., Bays, P. M., & Dayan, P. (2015). A probabilistic palimpsest model of visual short-term memory. *PLoS computational biology*, *11*(1), e1004003.

Musslick, S., & Cohen, J. D. (2021). Rationalizing constraints on the capacity for cognitive control. Trends in Cognitive Sciences, 25(9), 757-775.

Nili, H., Wingfield, C., Walther, A., Su, L., Marslen-Wilson, W., & Kriegeskorte, N. (2014). A toolbox for representational similarity analysis. *PLoS computational biology*, *10*(4), e1003553.

Nogueira, R., Rodgers, C. C., Bruno, R. M., & Fusi, S. (2021). The geometry of cortical representations of touch in rodents. *bioRxiv*, 2021-02.

Padoa-Schioppa, C., & Assad, J. A. (2006). Neurons in the orbitofrontal cortex encode economic value. Nature, 441(7090), 223-226.

Padoa-Schioppa, C., & Assad, J. A. (2008). The representation of economic value in the orbitofrontal cortex is invariant for changes of menu. Nature neuroscience, 11(1), 95-102.

Parthasarathy, A., Herikstad, R., Bong, J. H., Medina, F. S., Libedinsky, C., & Yen, S. C. (2017). Mixed selectivity morphs population codes in prefrontal cortex. Nature neuroscience, 20(12), 1770-1779.

Pedregosa, F., Varoquaux, G., Gramfort, A., Michel, V., Thirion, B., Grisel, O., Blondel, M., Prettenhofer, P., Weiss, R., Dubourg, V. and Vanderplas, J. (2011). Scikit-learn: Machine learning in Python. Journal of Machine Learning Research, 12, 2825-2830.

Plate, T. A. (1995). Holographic reduced representations. IEEE Transactions on Neural networks, 6(3), 623-641.

Pouget, A., & Sejnowski, T. J. (1997). Spatial transformations in the parietal cortex using basis functions. Journal of cognitive neuroscience, 9(2), 222-237.

Pu, S., Dang, W., Qi, X., & Constantinidis, C. (2022). Prefrontal neuronal dynamics in the absence of task execution. bioRxiv, 2022-09.

Rangel, A., Camerer, C., & Montague, P. R. (2008). A framework for studying the neurobiology of value-based decision making. Nature reviews neuroscience, 9(7), 545-556.

Raposo, D., Kaufman, M. T., & Churchland, A. K. (2014). A category-free neural population supports evolving demands during decision-making. Nature neuroscience, 17(12), 1784-1792.

Renart, A., & Machens, C. K. (2014). Variability in neural activity and behavior. Current opinion in neurobiology, 25, 211-220.

Rigotti, M., Barak, O., Warden, M. R., Wang, X. J., Daw, N. D., Miller, E. K., & Fusi, S. (2013). The importance of mixed selectivity in complex cognitive tasks. Nature, 497(7451), 585-590.

Roelfsema, P. R. (2023). Solving the binding problem: Assemblies form when neurons enhance their firing rate—they don't need to oscillate or synchronize. Neuron, 111(7), 1003-1019.

Roesch, M. R., Singh, T., Brown, P. L., Mullins, S. E., & Schoenbaum, G. (2009). Ventral striatal neurons encode the value of the chosen action in rats deciding between differently delayed or sized rewards. The Journal of neuroscience : the official journal of the Society for Neuroscience, 29(42), 13365–13376.

Roesch, M. R., Taylor, A. R., & Schoenbaum, G. (2006). Encoding of time-discounted rewards in orbitofrontal cortex is independent of value representation. Neuron, 51(4), 509-520.

Salinas, E., & Sejnowski, T. J. (2001). Book review: gain modulation in the central nervous system: where behavior, neurophysiology, and computation meet. The Neuroscientist, 7(5), 430-440.

Samejima, K., Ueda, Y., Doya, K., & Kimura, M. (2005). Representation of action-specific reward values in the striatum. Science, 310(5752), 1337-1340.

Saxena, S., & Cunningham, J. P. (2019). Towards the neural population doctrine. Current opinion in neurobiology, 55, 103-111.

Shadlen, M. N., & Movshon, J. A. (1999). Synchrony unbound: a critical evaluation of the temporal binding hypothesis. Neuron, 24(1), 67-77.

Sheahan, H., Luyckx, F., Nelli, S., Teupe, C., & Summerfield, C. (2021). Neural state space alignment for magnitude generalization in humans and recurrent networks. Neuron, 109(7), 1214-1226.

Smolensky, P. (1990). Tensor product variable binding and the representation of symbolic structures in connectionist systems. Artificial intelligence, 46(1-2), 159-216.

Sohn, H., Narain, D., Meirhaeghe, N., & Jazayeri, M. (2019). Bayesian computation through cortical latent dynamics. Neuron, 103(5), 934-947.

Sorscher, B., Ganguli, S., & Sompolinsky, H. (2022). Neural representational geometry underlies few-shot concept learning. *Proceedings of the National Academy of Sciences*, *119*(43), e2200800119.

Stoet, G., & Hommel, B. (1999). Action planning and the temporal binding of response codes. Journal of Experimental Psychology: Human Perception and Performance, 25(6), 1625–1640

Strait, C. E., Sleezer, B. J., Blanchard, T. C., Azab, H., Castagno, M. D., & Hayden, B. Y. (2016). Neuronal selectivity for spatial positions of offers and choices in five reward regions. Journal of neurophysiology, 115(3), 1098-1111.

Strait, C. E., Blanchard, T. C., & Hayden, B. Y. (2014). Reward value comparison via mutual inhibition in ventromedial prefrontal cortex. Neuron, 82(6), 1357-1366.

Strait, C. E., Sleezer, B. J., & Hayden, B. Y. (2015). Signatures of value comparison in ventral striatum neurons. PLoS Biol, 13(6), e1002173.

Sutherland, R. J., & Hoesing, J. M. (1993) Posterior cingulate cortex and spatial memory: A microlimnology analysis. In Neurobiology of cingulate cortex and limbic thalamus: A comprehensive handbook, pp. 461-477. Boston, MA: Birkhäuser Boston, 1993.

Tang, C., Herikstad, R., Parthasarathy, A., Libedinsky, C., & Yen, S. C. (2020). Minimally

dependent activity subspaces for working memory and motor preparation in the lateral prefrontal cortex. Elife, 9, e58154.

Treisman, A. M., & Gelade, G. (1980). A feature-integration theory of attention. Cognitive psychology, 12(1), 97-136.

Tsujimoto, S., Genovesio, A., & Wise, S. P. (2009). Monkey orbitofrontal cortex encodes response choices near feedback time. Journal of Neuroscience, 29(8), 2569-2574.

Urai, A. E., Doiron, B., Leifer, A. M., & Churchland, A. K. (2022). Large-scale neural recordings call for new insights to link brain and behavior. Nature neuroscience, 25(1), 11–19.

van Steenkiste, S., Locatello, F., Schmidhuber, J., & Bachem, O. (2019). Are disentangled representations helpful for abstract visual reasoning? Advances in Neural Information Processing Systems, 32.

von der Malsburg, C. (1999). The what and why of binding: the modeler's perspective. Neuron, 24(1), 95-104.

Vyas, S., Golub, M. D., Sussillo, D., & Shenoy, K. V. (2020). Computation through neural population dynamics. Annual Review of Neuroscience, 43, 249-275.

Watanabe, S. (2013). A widely applicable Bayesian information criterion. Journal of Machine Learning Research, 14, 867-897.

Widge, A. S., Heilbronner, S. R., & Hayden, B. Y. (2019). Prefrontal cortex and cognitive control: new insights from human electrophysiology. F1000Research, 8.

Wunderlich, K., Rangel, A., & O'Doherty, J. P. (2009). Neural computations underlying action-based decision making in the human brain. Proceedings of the National Academy of Sciences, 106(40), 17199-17204.

Xie, Y., Hu, P., Li, J., Chen, J., Song, W., Wang, X. J., ... & Wang, L. (2022). Geometry of sequence working memory in macaque prefrontal cortex. Science, 375(6581), 632-639.

Yao, Y., Vehtari, A., Simpson, D., Gelman A. (2017). Using stacking to average Bayesian predictive distributions. arXiv preprint arXiv:1704.02030v3

Yim, M. Y., Cai, X., & Wang, X. J. (2019). Transforming the choice outcome to an action plan in monkey lateral prefrontal cortex: A neural circuit model. Neuron, 103(3), 520-532.

Yoo, S. B. M., & Hayden, B. Y. (2020). The transition from evaluation to selection involves neural subspace reorganization in core reward regions. Neuron, 105(4), 712-724.

Yoo, S. B. M., & Hayden, B. Y. (2018). Economic choice as an untangling of options into actions. Neuron, 99(3), 434-447.

Yoo, S. B. M., Sleezer, B. J., & Hayden, B. Y. (2018). Robust encoding of spatial information in orbitofrontal cortex and striatum. Journal of cognitive neuroscience, 30(6), 898-913.

Yoo, S. B. M., Tu, J. C., Piantadosi, S. T., & Hayden, B. Y. (2020). The neural basis of predictive pursuit. Nature neuroscience, 23(2), 252-259.

Yoo, S. B. M., Hayden, B. Y., & Pearson, J. M. (2021). Continuous decisions. Philosophical Transactions of the Royal Society B, 376(1819), 20190664.

Zhang, J., & Abbott, L. F. (2000). Gain modulation of recurrent networks. Neurocomputing, 32, 623-628.

# Supplementary Figures

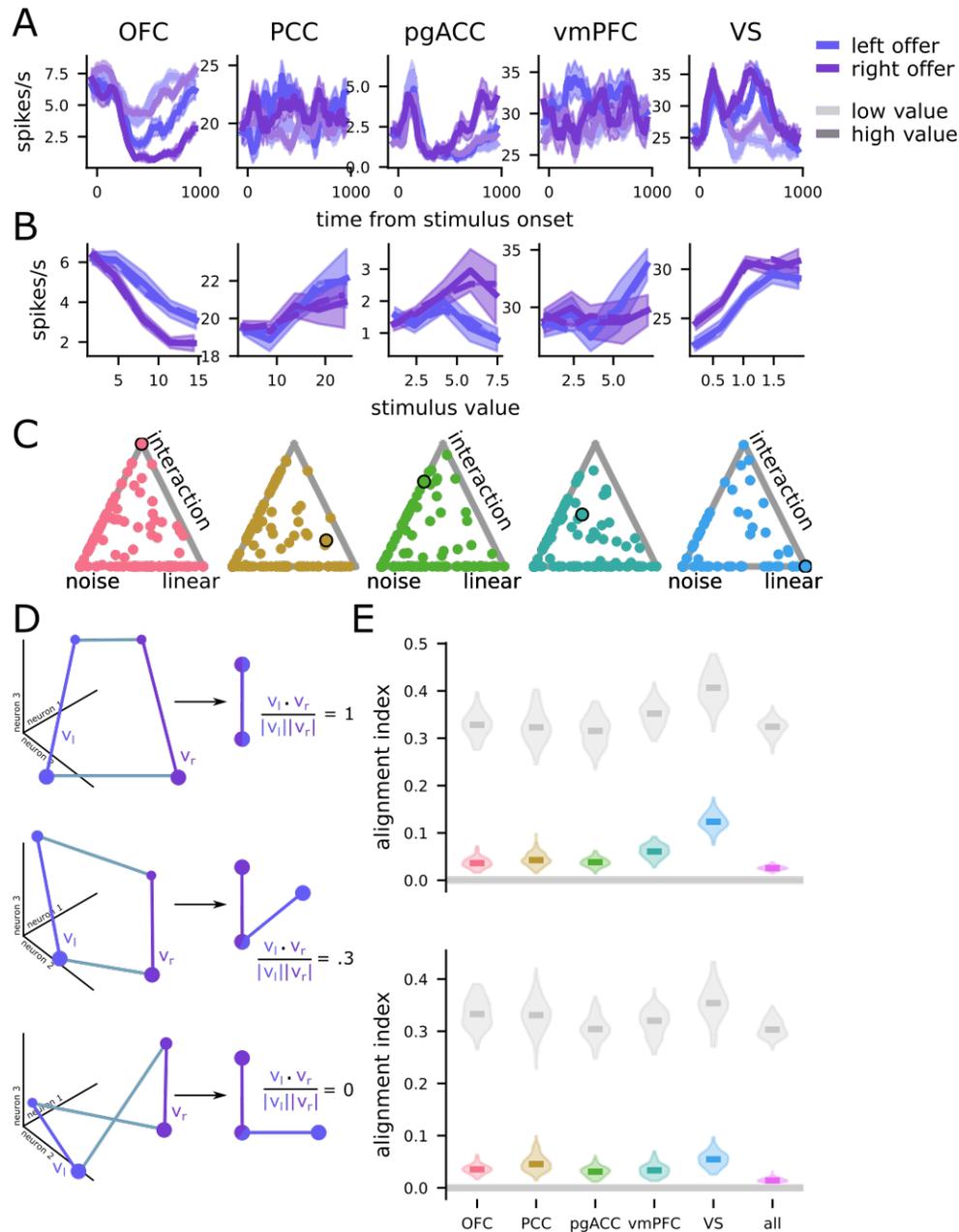

**Figure S1.** Example neurons, model comparison, and subspace correlations for the linear models with a nonlinear value representation. **A.** The firing rates of example neurons from each region during the offer window, shown for high and low value offers presented on the left or right side (100 ms boxcar filter, shaded area is SEM). **B.** The value-response function for each neuron in **A**. The value-response function fit by the linear regression model with a b-spline value encoding and an interaction term is overlaid (dashed lines). The b-spline representation uses 4 knots and is degree 2. **C.** A simplex showing the weight given to each of the noise-only, linear, and interaction regression models by the Bayesian model stacking analysis. The points corresponding

to the example neurons shown in **A** and **B** have dark outlines here. Both the linear and interaction categories include both linear and spline value representation models. **D.** Schematic of three different representational geometries that would lead to different subspace correlation results. (top) Two perfectly aligned value vectors $v_l$ and $v_r$ in population space (left) would produce a subspace correlation close to 1 (right). (middle) Two partially aligned value vectors $v_l$ and $v_r$ in would produce a subspace correlation between 0 and 1 (note there is an additional possibility: partially aligned but negatively correlated subspaces; not schematized). (bottom) Two unaligned value vectors $v_l$ and $v_r$ would produce a subspace correlation close to 0. **E.** Alignment indices for all regions for the offer presentation window. The gray point is the subspace correlation expected if the left- and right value subspaces were aligned and corrupted only due to noise. **E.** Same as **D.** for the delay period.

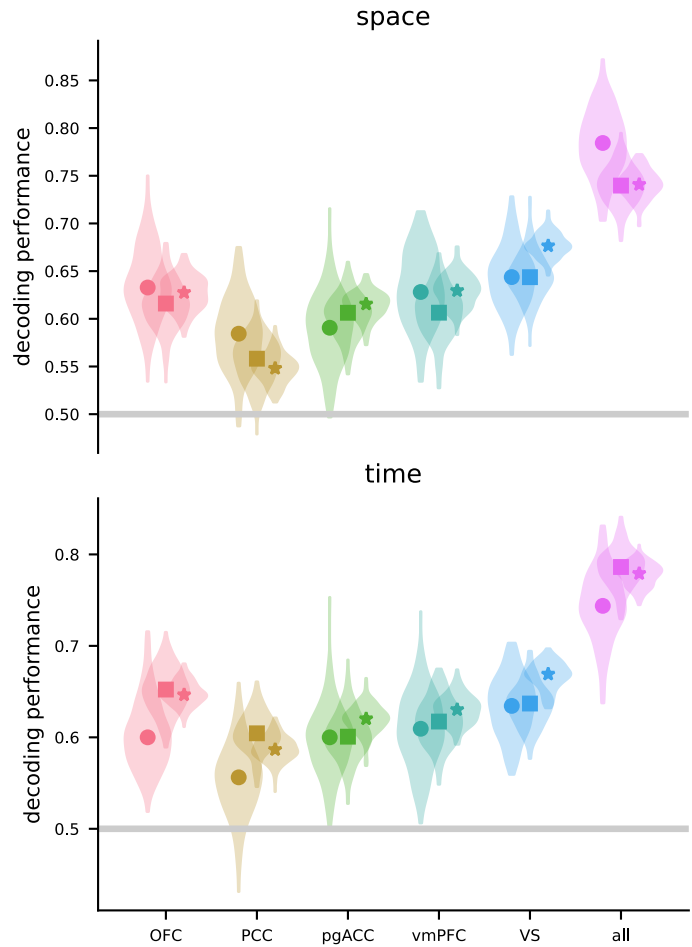

**Figure S2.** Value decoding, value generalization, and predicted value generalization of the code within each recorded region. **A.** Pseudopopulation value decoding performance (circles), generalization performance (squares, trained on offers from one side, tested on offers from the other side), and predicted generalization performance (stars) shown for each region and the neural population combined across regions ("all"), shown for the left and right value comparison. The violin plot shows the values produced from two hundred bootstrap resamples of the trials. **B.** The same as **A** except shown for the offer 1 and offer 2 comparison. PCC did not have the required number of trials (160) for each condition.

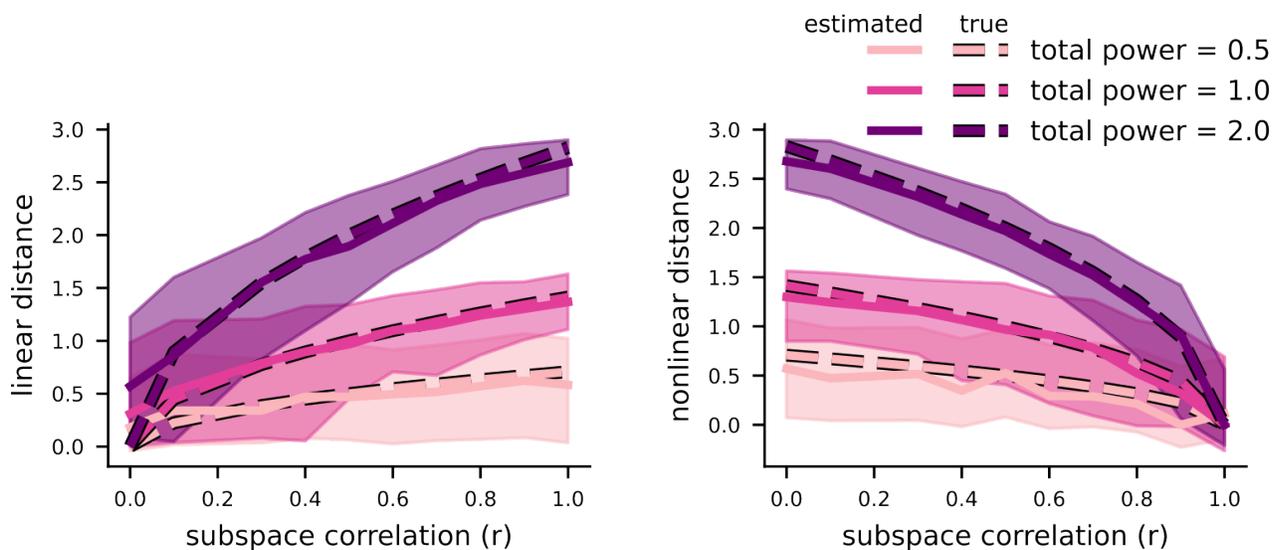

**Figure S3.** Accurate recovery of linear and nonlinear distances from simulated data. The true (dashed line) and estimated (solid line with error bars) linear (left) and nonlinear (right) distances from simulated data. Our decomposition accurately recovers the linear and nonlinear distances, as the true value is always within error bars of the estimated value and the estimate is typically unbiased.

**Neurons with heterogeneous representations of value across positions drive subspace separation**

What is the single neuron basis of the subspaces? One simple possibility is that the subspaces may be composed almost entirely of nonlinearly selective neurons that each only represent the value of offers at a single spatial position (i.e., neurons with canonical spatial receptive fields). The full population, then, would be composed of two largely separate subpopulations: one for offers on the left and another for offers on the right (**Figure S4A**, left, and S**4B**, top). This organization could also be viewed as a gain modulation code, where spatial position strongly modulates the value code of single neurons, without changing their tuning (e.g., **Figure S4E**, left: a left offer tuned cell). Alternatively, subspace separation could be achieved by neurons with heterogeneous nonlinear responses to offer value and position, and that contribute activity to multiple subspaces (e.g., **Figure S4E**, right; Fusi et al., 2016; Tang et al., 2020). We refer to this as the *shared population hypothesis* (**Figure S4A**, right, and S**4B, bottom**).

These different coding strategies lead to distinct predictions across the population. If the population was dominated by cells that respond more strongly to offers presented at one of the two positions (or even cells that respond only to offers presented at one position), then we would expect a bimodal distribution of differences in selectivity for left and right value across the population (**Figure S4C;** Elsayed et al., 2016). Alternatively, if the population is composed of neurons that represent value differently across the two positions, but with similar strength, then we would expect this distribution to be unimodal (**Figure S4C**). We discriminated between these two possibilities using the regression coefficients to characterize each neuron's value representation for each location. We then asked if the distributions of firing rate differences diverged from a unimodal distribution using Hartigan's dip test. We then repeated this analysis for all time-windows. We found no evidence for bimodality ($p > 0.9$ for all areas and subjects). The first offer on time window provides an example of the typical distribution (**Figure S4D**). This result therefore supports the idea that the value of left and right offers are encoded in distinct subspaces, but not in distinct populations of neurons (as would be expected by a simple spatial receptive field model). Note that this finding of non-categoricality is consistent with several other recent studies emphasizing non-categorical neural responses (Blanchard et al., 2018; Raposo et al., 2014; Kaufman et al., 2022).

The lack of bimodality implies there may be several types of nonlinear encoding neurons that drive the subspaces. Next, we show the left and right subspaces are supported by neurons with both spatially-tuned gain modulation and heterogenous value representations (e.g., **Figure S4E**). Specifically, we quantified the proportion of each neuron type in each brain region. First, we searched for neurons that contributed to both a left and right subspace within an analysis epoch, using a measure of subspace contribution (Xie et al., 2022). The subspace contribution effectively computes the proportion of variance a neuron contributes to a subspace. Next, remaining neurons were classified based on whether their preferred value was either constant (gain modulated) or shifted across subspaces (heterogeneous nonlinear). Examining the response profiles of neurons meeting the multi-subspace criterion shows both those with gain modulation and those with shifting tuning to value that depends on the subspace (**Figure 4E**). Averaging across all regions and windows, we found that both types were found in roughly equal proportions. These results rule out simple spatial subpopulation or wide-spread gain modulation explanations of value subspaces and support the idea that the two subspaces are represented in the same population of neurons, but semi-orthogonal axes in population space.

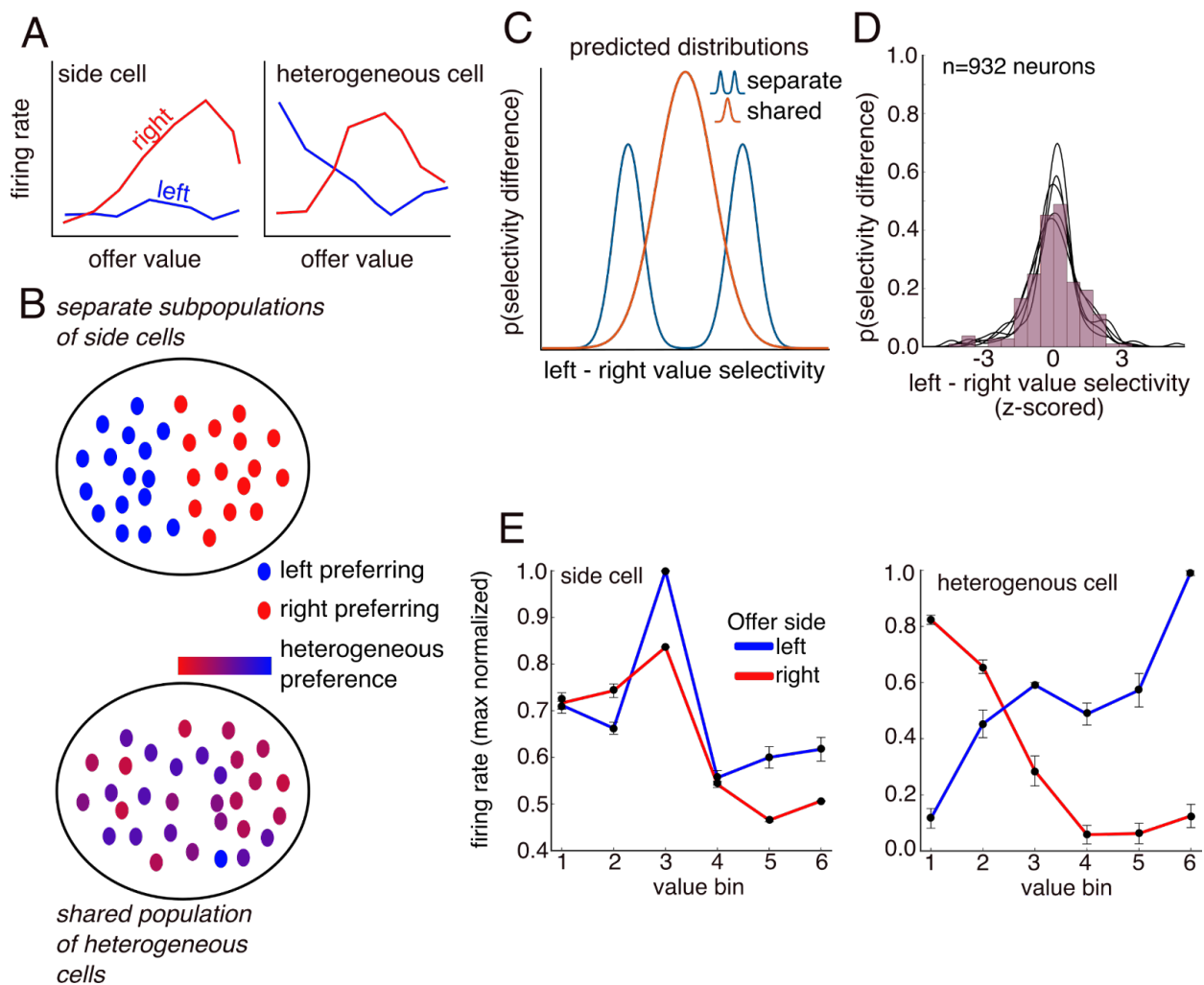

**Figure 4.** Understanding the nonlinear selectivity underlying subspaces and binding. **A.** Two schematic kinds of nonlinear selectivity: (left) A strong side preference and (right) two different response profiles to the two different sides. **B.** The two kinds of selectivity give rise to two distinct hypotheses for selectivity across the population: (top) Separate subpopulations, each composed of cells with a strong preference for one of the two positions and (bottom) a single population composed of cells with heterogeneous response profiles for the two sides. Both forms of nonlinear population selectivity achieve subspace binding. (**C**). These hypotheses (**A-B**) predict differences in how the distribution of selectivity differences for left and right value subspaces will appear. The separate subpopulations hypothesis predicts a closer to bimodal distribution (blue line), while the shared, heterogeneous hypothesis predicts a unimodal distribution (orange line). (**D**) estimated distribution of differences in value selectivity for left and right subspaces for offer 1 on time window. Each line is a different region, showing they are all unimodal. (**E**). Example OFC nonlinear encoding neurons showing: one has a weak side preference (left) and the other has a heterogeneous response profile (right).